# N-, B-, P-, Al-, As-, Ga-graphdiyne/graphyne lattices: First-principles investigation of mechanical, optical and electronic properties

Bohayra Mortazavi[*,b,c], Masoud Shahrokhi[d], Mohamed E. Madjet[e], Tanveer Hussain[f,g], Xiaoying Zhuang[c] and Timon Rabczuk[#,a]

[a]*Institute of Research & Development, Duy Tan University, Quang Trung, Da Nang, Vietnam.*
[b]*Institute of Structural Mechanics, Bauhaus-Universität Weimar, Marienstr. 15, D-99423 Weimar, Germany.*
[c]*Cluster of Excellence PhoenixD (Photonics, Optics, and Engineering–Innovation Across Disciplines), Hannover, Germany.*
[d]*Department of Physics, Faculty of Science, Razi University, Kermanshah, Iran.*
[e]*Qatar Environment and Energy Research Institute, Hamad Bin Khalifa University, Doha, Qatar.*
[f]*School of Molecular Sciences, The University of Western Australia, Perth, WA 6009, Australia. Centre for*
[g]*Theoretical and Computational Molecular Science, Australian Institute for Bioengineering and Nanotechnology, The University of Queensland, Brisbane, Queensland 4072, Australia*

**Abstract**

Graphdiyne and graphyne are carbon-based two-dimensional (2D) porous atomic lattices, with outstanding physics and excellent application prospects for advanced technologies, like nanoelectronics and energy storage systems. During the last year, B- and N-graphdiyne nanomembranes were experimentally realized. Motivated by the latest experimental advances, in this work we predicted novel N-, B-, P-, Al-, As-, Ga-graphdiyne/graphyne 2D lattices. We then conducted density functional theory simulations to obtain the energy minimized structures and explore the mechanical, thermal stability, electronic and optical characteristics of these novel porous nanosheets. Acquired theoretical results reveal that the predicted carbon-based lattices are thermally stable. It was moreover found that these novel 2D nanostructures can exhibit remarkably high tensile strengths or stretchability. The electronic structure analysis reveals semiconducting electronic character for the predicted monolayers. Moreover, the optical results indicate that the first absorption peaks of the imaginary part of the dielectric function for these novel porous lattices along the in-plane directions are in the visible, IR and near-IR (NIR) range of light. This work highlights the outstanding properties of graphdiyne/graphyne lattices and recommends them as promising candidates to design stretchable energy storage and nanoelectronics systems.
Corresponding authors: *bohayra.mortazavi@gmail.com, #timon.rabczuk@uni-weimar.de;

# 1. Introduction

Two-dimensional (2D) materials can be currently considered as the most appealing class of materials, stemming from their application prospects in critical technologies, such as the; nanoelectronics, nanooptics, bio- and nano-sensors, mobile energy storage/conversion systems and structural components in aerospace. The importance of 2D materials came into consideration originally after the rise of graphene [1,2], which exhibits uniquely high mechanical properties [3] and thermal conductivity [4,5] and highly attractive optical and electronic properties [6–9]. The great success of graphene motivated the design and synthesis of different classes of 2D materials. Despite of exceptional properties of graphene, it also exhibits some limitations for particular applications. These limitations of graphene have been acting as great motivations to search, predict, design and fabricate novel 2D materials. For example, pristine graphene does not show an electronic band-gap, which has encouraged the synthesis of novel semiconducting 2D lattices, like: transition metal dichalcogenides such as the $MoS_2$ and $WS_2$ [10–12], phosphorene [13,14] and 2D carbon nitrides [15–20]. As another example, densely packed atomic lattice of defect-free graphene with all $sp^2$ carbon atoms, lead to moderate charge capacities for the application as anode materials for rechargeable metal-ion batteries. In this case, other 2D materials with porous atomic structures and hybrid sp and $sp^2$ carbon atoms, like graphdiyne [21,22] nanosheets can yield considerably higher charge capacities than graphene for the application as anode electrodes in various rechargeable metal-ion battery technologies [23–27].

Among the various classes of 2D materials, graphdiyne/graphyne full carbon and porous lattices, can serve as promising complementary materials to graphene. The interests toward these class of materials, date back to a couple of decades before the rise of graphene. In this regard, Baughman *et. al* [21] in 1987 predicted numerous graphdiyne/graphyne atomic lattices. Interestingly, on the basis of first-principles calculations it was known that different graphdiyne family members can exhibit semiconducting electronic character [26,28]. Densely packed atomic lattice of graphene limits its stretchability and results in brittle failure mechanism. On the other side, porous atomic structures of graphdiyne nanosheets can facilitate the deformation upon the mechanical loading, which can help to design foldable and stretchable devices [29,30]. The porous lattices of graphdiyne/graphyne nanosheets also provide optimum conditions for the access to the active sites in these materials, favourable for the adatoms adsorption and diffusion, highly desirable for the design of next generation rechargeable metal-ion batteries [31–33], hydrogen storage systems [34,35] and catalysts [36]. Unlike the graphene, non-uniform sp and $sp^2$ carbon atomic types along with porous atomic structures of graphdiyne nanosheets, scatter the phonons and accordingly substantially suppress the thermal conductivity, which can suggest them as promising candidates to design carbon-based thermoelectric devices [37,38].

Despite of outstanding application prospects of graphdiyne/graphyne nanomaterials, their experimental realization remained a critical challenge till 2010, when for the first time a graphdiyne structure was fabricated by Li *et al.* [22], employing a cross-coupling reaction using the hexaethynylbenzenethe. In 2017, Matsuoka *et al.* [39] reported the experimental realization of two different graphdiyne structures at a gas/liquid or liquid/liquid interface. The method devised by Matsuoka *et al.* [39] has been used as a promising platform to fabricate novel graphdiyne lattices. In a very recent advance, Kan *et al.* [40] fabricated nitrogen-graphdiyne nanomembranes. Wang *et al.* [41] also most recently reported the synthesis of boron-graphdiyne nanosheets. These latest advances [40,41] highlight the possibility of the design of novel graphdiyne nanomaterials, in which the $sp^2$ carbon atoms in the original graphdiyne lattices can be replaced with nitrogen or boron atoms.

Recent and continuous experimental advances concerning the fabrication of full-carbon and N- and B-graphdiyne [22,39–41] nanosheets, reveal substantial steps toward the practical applications of these nanomaterials in various advanced and critical technologies. In our previous works, we found that N-graphdiyne [42] and B-graphdiyne [25] nanosheets show semiconducting electronic character, high stretchability, low thermal conductivity and very promising optical properties. Our density functional theory calculations have also confirmed that, B-graphdiyne [25], N-graphdiyne [43] and N-, P-triphenylene-graphdiyne [44] can yield ultrahigh charge capacities as anode materials for different metal-ion battery technologies. Successful synthesis of these novel nanomaterials subsequently emphasizes the importance of new studies to provide an in-depth understanding of their intrinsic properties. In addition, these advances raise a simple question concerning the stability, intrinsic properties and possibility of the synthesize of other graphdiyne/graphyne lattices. Because of the complexities of experimental fabrication and characterization techniques, theoretical studies can be considered as the fastest approaches to examine new compositions, estimate their properties, find potential applications and suggest possible synthesis routes [45–47]. In this work we accordingly predicted novel N-, B-, P-, Al-, As-, Ga-graphdiyne/graphyne nanosheets. In order to provide insight into the thermal stability, mechanical properties and electronic/optical properties of these nanomembranes, we conducted extensive density functional theory simulations. The results acquired by the first-principles simulations confirm the stability and reveal highly desirable properties of this novel class of porous 2D materials, which may serve effectively for the next generation nanoelectronics and energy storage systems.

## 2. Computational methods

Density functional theory (DFT) calculations have been performed employing the *Vienna Ab-initio Simulation Package* (VASP) [48–50]. The DFT calculations are within the frame work of generalized gradient approximation (GGA) in the form of Perdew–Burke–Ernzerhof (PBE) [51] for the exchange correlation potential and the ion–electron interaction is treated using the projector augmented wave (PAW) [52] method. We used a plane-wave cutoff energy of 500 eV and the convergence criteria for the electronic self consistence-loop was set to be $10^{-4}$ eV. VESTA [53] package was used for the illustration of atomic structures and charge densities as well. Energy minimized structures were acquired by altering the size of the unit-cells and then employing the conjugate gradient method for the geometry optimizations. The convergence criteria for the Hellmann–Feynman forces on each atom was taken to be 0.01 eV/Å. Mechanical properties were evaluated by performing uniaxial tensile simulations [54]. Since the PBE/GGA underestimates the band-gap values, we used the screened hybrid functional HSE06 [55] to provide more accurate estimations. For the Brillouin zone integration, a 10×10×1 Γ centered Monkhorst-Pack [56] k-point mesh was used to ensure converged electronic results within PBE and HSE06. The optical properties were estimated using the random phase approximation (RPA) constructed over PBE results. For the simulations involving the optical properties, Brillouin zone was sampled by 14×14×1 grids. A detailed explanation of optical properties calculations was described in our previous works [57–59]. Thermal stability was examined by conducting the ab-initio molecular dynamics (AIMD) simulations for the energy minimized unit-cells, using the Langevin thermostat with a time step of 1 fs and 2×2×1 Monkhorst-Pack k-point mesh size.

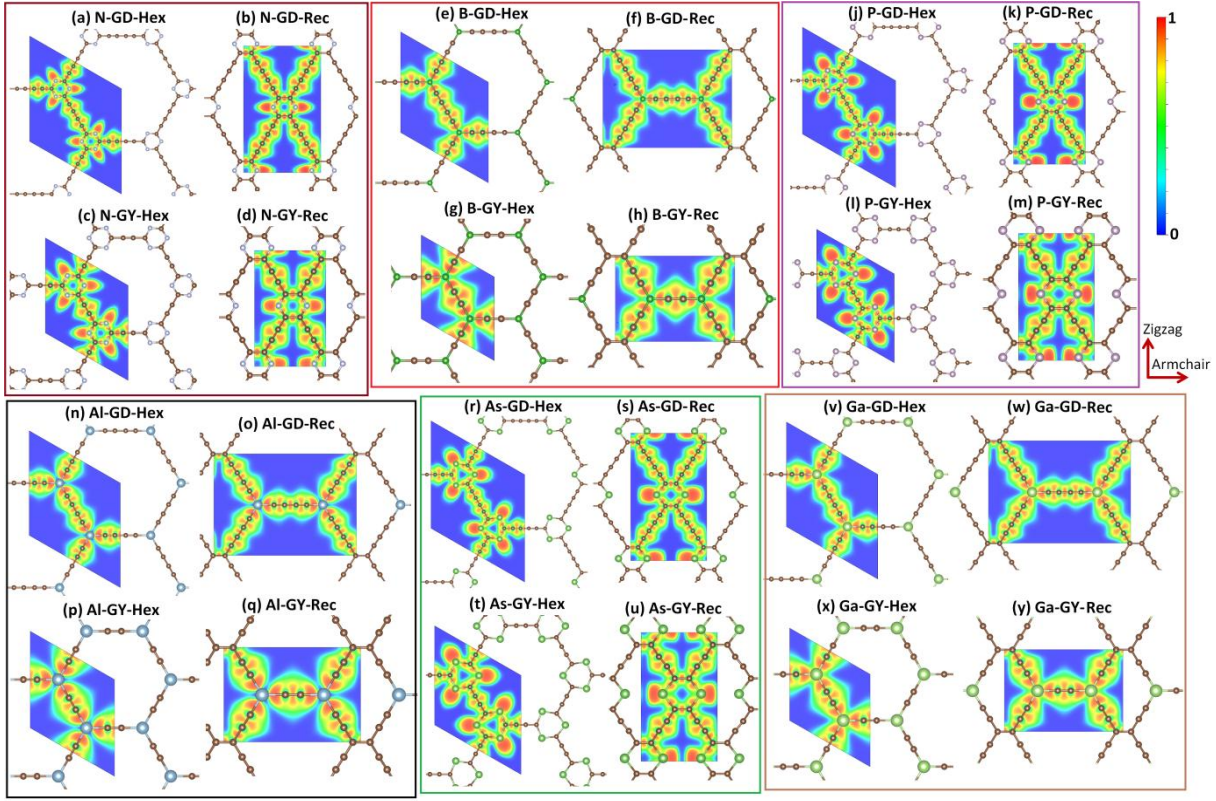

**Fig. 1,** Atomic structures of N-, B-, P-, Al-, As-, Ga-graphdiyne (GD) or -graphyne (GY) monolayers. Contours illustrate the electron localization function [61] within the unit-cell.

## 3. Results and discussions

In Fig.1, energy minimized N-, B-, P-, Al-, As-, Ga-graphdiyne/graphyne monolayers with graphene-like hexagonal (Hex) and rectangular (Rec) atomic lattices are illustrated. Constructed graphdiyne and graphyne nanosheets show similar atomic structures, however, the number of carbon atoms in the carbon-carbon connecting links are different. In the cases of graphyne (GR) and graphdiyne (GD) lattices, carbon links are build from 1 (2 individual C atoms) and 2 (4 individual C atoms) carbon-carbon triple bonds, respectively. To the best of our knowledge, full carbon graphyne nanosheets have not been synthesized up to now. Among the studied monolayers, only N-graphdiyne (shown in Fig. 1a and 1b) and hexagonal B-graphdiyne (shown in Fig. 1e) nanosheets have been experimentally realized by Kan *et al.* [40] and Wang *et al.* [41], respectively. As it was originally explored by Baughman *et al.* [21], graphyne/graphdiyne lattices may show numerous structures. In this way, in order to limit the extent of our study, we only considered two particular hexagonal and rectangular lattices for each substitutive element, as depicted in Fig. 1. In addition, for the substitution of carbon atoms in the original graphyne/graphdiyne lattices, numerous configurations can be taken into considerations. In this work, for the P- and As-graphyne/graphdiyne structures, we only considered the replacement of the N atoms with P and As counterparts in the experimentally realized N-graphdiyne [40] sheets. Such a substitution mechanism is also in accordance with other synthesized carbon-nitride 2D materials, like; triazine-based graphitic carbon nitrides [15] and nitrogenated holey graphene [60]. A similar strategy was also considered as the basis for the construction of Al- and Ga-graphyne/graphdiyne lattices, in which the B atoms in the experimentally realized B-graphdiyne [41] are substituted with Al and Ga atoms. On this experimentally stemmed basis, the connecting links in the all graphyne/graphdiyne lattices are kept to be made entirely from the carbon atoms. More importantly, the crystallinity of all predicted nanosheets are fully preserved, meaning that the first-principles simulations over the unit-cells can provide accurate estimations.

In analogy to graphene and in order to examine the anisotropy in the mechanical response, for every monolayer armchair and zigzag directions were defined (as depicted in Fig. 1). To briefly analyse the bonding nature in these novel 2D systems, in Fig. 1 the electron localization function (ELF) [61] within the unit-cells is illustrated, which is a spatial function and takes a value between 0 and 1. As it is clear, electron localization occurs around the center of all bonds in these nanosheets, confirming the covalent bonding. It is worthy to note that among the considered atoms for the functionalization of graphyne/graphdiyne lattices, only the nitrogen atom exhibit a higher electronegativity than carbon. Such that as shown in Fig. 1, only for the N-graphyne/graphdiyne lattices the electron localization on the N-C bonds is extended toward the nitrogen atoms and in the rest of the cases along the heteronuclear bonds the electron localization is extended toward the carbon atoms. Worthy to note that the unit-cells of energy minimized monolayers are all provided in the supplementary information document. In Table 1, the lattice constants of energy minimized structures and some important bond lengths are summarized. As it is clear, the maximum changes in the bond lengths are happening in the cases of carbon atoms connected to the functionalizing atoms. Interestingly, the bond lengths along the carbon-carbon links are almost unaffected for different functionalizing atoms and the maximum difference along the all constructed monolayers is less than 3%. In addition, for a particular substituting atom the corresponding bond lengths in different lattices are very close.

**Table 1**, Lattice parameters of energy minimized graphyne/graphdiyne monolayers. $L_{X-C}$, $L_{X\equiv C}$, $L_{CX-C}$ and $E_{uc}$ are, respectively, the bond lengths of C-X (X=N, B, P, Al, As, Ga), C-C triple bonds and XC-C, and energy per atom of a unit-cell. For the structures with rectangular lattices, the lattice constants are given along the armchair and zigzag directions. All the energy minimized unit-cells are given in the supplementary information document

| Structure | Lattice constant (Å) | $L_{X-C}$(Å) | $L_{C\equiv C}$(Å) | $L_{XC-C}$(Å) | $E_{uc}$(eV/atom) |
|---|---|---|---|---|---|
| N-GD-Hex | 16.037 | 1.350 | 1.224 | 1.424 | -8.30 |
| N-GD-Rec | 9.670$^{arm.}$, 15.970$^{zig.}$ | 1.345 | 1.227 | 1.407 | -8.36 |
| N-GY-Hex | 11.580 | 1.348 | 1.215 | 1.429 | -8.33 |
| N-GY-Rec | 7.052$^{arm.}$, 11.545$^{zig.}$ | 1.344 | 1.222 | 1.412 | -8.42 |
| B-GD-Hex | 11.847 | 1.512 | 1.234 | | -6.22 |
| B-GD-Rec | 15.118$^{arm.}$, 11.561$^{zig.}$ | 1.51 | 1.236 | | -8.12 |
| B-GY-Hex | 7.400 | 1.520 | 1.231 | | -7.80 |
| B-GY-Rec | 9.912$^{arm.}$, 7.146$^{zig.}$ | 1.52 | 1.231 | | -8.06 |
| P-GD-Hex | 17.076 | 1.765 | 1.235 | 1.399 | -6.91 |
| P-GD-Rec | 9.902$^{arm.}$, 16.757$^{zig.}$ | 1.763 | 1.235 | 1.399 | -7.89 |
| P-GY-Hex | 12.623 | 1.763 | 1.230 | 1.404 | -7.24 |
| P-GY-Rec | 7.314$^{arm.}$, 12.352$^{zig.}$ | 1.763 | 1.231 | 1.405 | -7.75 |
| Al-GD-Hex | 13.191 | 1.896 | 1.235 | | -5.84 |
| Al-GD-Rec | 16.691$^{arm.}$, 11.980$^{zig.}$ | 1.896 | 1.235 | | -7.83 |
| Al-GY-Hex | 8.739 | 1.90 | 1.234 | | -6.97 |
| Al-GY-Rec | 11.153$^{arm.}$, 7.769$^{zig.}$ | 1.90 | 1.234 | | -7.57 |
| As-GD-Hex | 17.443 | 1.904 | 1.242 | 1.384 | -6.64 |
| As-GD-Rec | 10.092$^{arm.}$, 17.011$^{zig.}$ | 1.907 | 1.237 | 1.394 | -7.71 |
| As-GY-Hex | 12.994 | 1.902 | 1.239 | 1.388 | -6.85 |
| As-GY-Rec | 7.439$^{arm.}$, 12.665$^{zig.}$ | 1.909 | 1.235 | 1.399 | -7.51 |
| Ga-GD-Hex | 13.250 | 1.916 | 1.232 | | -5.63 |
| Ga-GD-Rec | 16.654$^{arm.}$, 12.068$^{zig.}$ | 1.914 | 1.232 | | -7.68 |
| Ga-GY-Hex | 8.795 | 1.924 | 1.229 | | -6.53 |
| Ga-GY-Rec | 11.202$^{arm.}$, 7.799$^{zig.}$ | 1.920 | 1.230 | | -7.32 |

Presenting good mechanical properties are critical for the practical application of a material in various devices, and these properties should be elaborately known as requirements for the engineering designs. We therefore first examine the mechanical responses of these novel carbon-based porous 2D lattices, by conducting uniaxial tensile simulations. In order to simulate the anisotropy in the mechanical responses of predicted 2D materials, the uniaxial tensile simulations were conducted along the armchair and zigzag directions. For the uniaxial tensile simulations, the periodic simulation box size along the loading direction was increased gradually. In order to observe the uniaxial stress-conditions, the simulation box size along the sheet perpendicular direction of loading was adjusted to reach a negligible stress [62–65]. After the adjustments of the simulation box sizes and accordingly rescaling the atomic positions, an energy minimization step within the conjugate gradient method was conducted to allow the rearrangement of atomic positions. In order to understand the deformation mechanism in considered novel 2D materials, in Fig. 2, the first-principles results for the uniaxial stress-strain responses of B- and N-graphyne/graphdiyne nanosheets elongated along the zigzag and armchair directions are compared. As it is clear, graphyne and graphdiyne lattices show distinctly different stress-strain relations. Principally for graphdiyne lattices, the predicted uniaxial stress-strain curves are not likely to the most of conventional materials, those the stress-strain curves exhibit initial linear responses, corresponding to the linear elasticity. The deformation of considered graphdiyne nanosheets are more likely to the elastomers and rubbers, in which the elastic modulus is not constant initially and increases by increasing the strain level. Results shown in Fig. 2 also reveal that for a considered graphyne/graphdiyne lattice, the stress-strain curves along the different loading directions are clearly different, which confirm anisotropic mechanical properties.

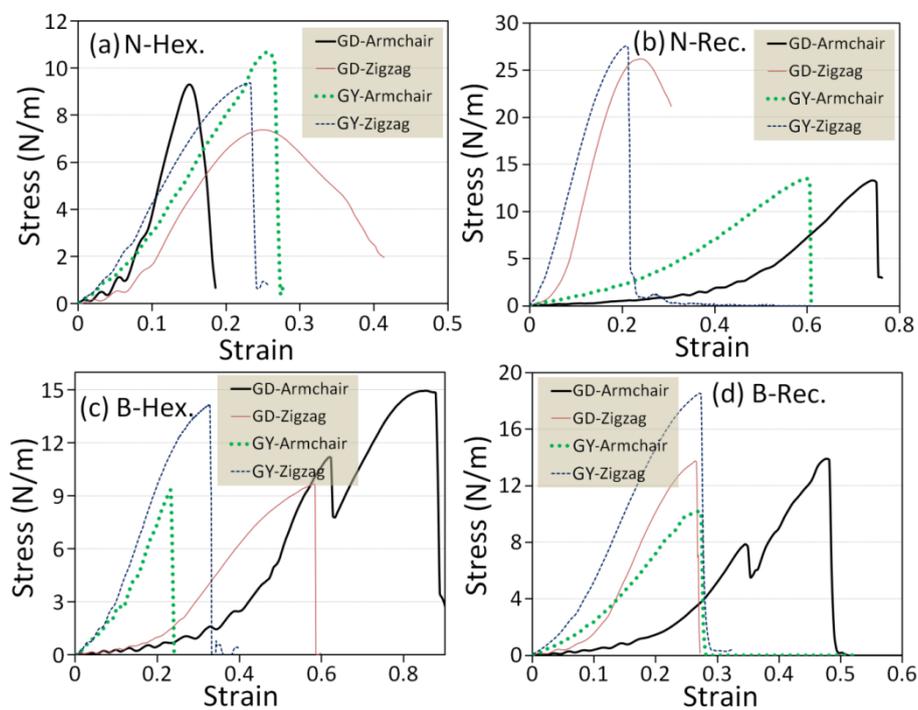

**Fig. 2**, Uniaxial stress-strain responses of single-layer N- and B-graphdiyne (GD) or -graphyne (GY) stretched along the armchair and zigzag directions.

The unusual mechanical response of graphyne/graphdiyne nanosheets can be mostly attributed to their porous atomic lattices and the configuration of full carbon chains with respect to the loading direction. Worthy to remind that for the most of materials with densely packed structures, like graphene, ceramics and metals, during the uniaxial loading the stretching of the structure can be achieved only by increasing the atomic distances, those oriented along the loading direction. Because of the harmonic nature of the bond stretching, the stress-strain curves usually start with linear relations. As it has been elaborately discussed in our previous works concerning the N-graphdiyne [42] and B-graphdiyne [25], in these nanomaterials at the initial strain levels the deformation is achieved mainly by the rotation of full carbon chains and subsequent alignment these chains toward the loading direction. In graphdiyne nanosheets, the bond stretching starts to dominate the deformation process at higher strain level, at which the bond rotation and structural deflection have already become difficult. Since these deformation mechanisms; bond stretching and bond rotation occur simultaneously, the stress-strain relations show irregular and noisy patterns. The regions that the bond stretching dominates the deformation can be realized by finding the points in which the stress values increase sharply. In contrast, the slight increases or drops in the stress values can reveal the occurrences of bond rotations or remarkable contraction of the sheet along the width. Results shown in Fig. 2 for the B-graphdiyne monolayers with both hexagonal and rectangular lattices, unexpected and sudden sharp drops are observable in the stress-strain relations. At these points, we found that due to sudden and considerable contractions of the sheets along the width (perpendicular direction of the loading), some parts of the stresses in the stretched bonds relieve, resulting in the overall declines of the effective stress values. In the cases of graphyne nanosheets, since their atomic lattices show a lower level of porosity in comparison with their graphdiyne counterparts, bond rotations and sudden contractions of the structures along the width of the sheets become more limited, resulting in the stress-strain curves more likely to the conventional ones, with more observable initial linear elasticity. As a drawback of the reducing of the porosity and the limitation of bond rotations in graphyne lattices, in these nanomembranes generally ruptures occur at earlier strain levels as compared with their graphdiyne counterparts, leading to more brittle failure mechanisms and more importantly declined stretchability. Nevertheless few exceptions can be yet found, as for example along the armchair direction the stretchability of N- graphyne was notably found to be around twice of that for the N-graphdiyne (Fig. 2a). The results shown in Fig. 2 reveal that the no general trend for tensile strengths of different graphyne/graphdiyne nanosheets can be established. One expects that due to the higher porosities of graphdiyne nanosheets they would show lower tensile strengths in comparison with their denser graphyne counterparts. Nevertheless, depending on the loading direction and atomic configurations, remarkable bond rotations and in-width contraction of graphdiyne lattices during the uniaxial loading can generally result in the increasing of the density and finally leading to comparable tensile strengths to their graphyne counterparts. With respect to the mechanical properties, nonetheless the main advantages of porous 2D materials like graphyne/graphdiyne lattices lie in their superior stretchability and flexibility, in comparison densely packed counterparts like graphene and $MoS_2$. Nonetheless, the tensile strengths of graphyne/graphdiyne lattices are still among the most critical parameters for the design of nanodevices. In Table 2 we accordingly summarize and compare the strain at failures (representative of stretchability) and tensile strengths of all studied graphyne/graphdiyne nanosheets loaded uniaxially along the armchair and zigzag directions. As it is clear, B-graphdiyne with the hexagonal atomic lattice exhibits the highest stretchability among the all studied porous nanosheets. Al- and Ga- graphyne/graphdiyne nanosheets show relatively low stretchability, only slightly higher than that of the graphene. Notably, some graphyne/graphdiyne nanosheets can exhibit remarkably high tensile strengths around 25 N/m, which are only ~35% lower than that of the graphene (~37-40 N/m [66]). Among the all studied nanosheets, the softest structure was found to be As-graphdiyne with the hexagonal atomic lattice, with a tensile strength of 3.9 N/m along the zigzag direction. As an interesting matter of fact, the tensile strength of As-graphdiyne is higher than that of the some densely packed 2D materials, like; stanene [67].

**Table 2**, Ultimate tensile strength and strain at tensile strength point (stretchability) of single-layer graphyne/graphdiyne lattices elongated along the armchair and zigzag directions.

| Structure | Strain at tensile strength point | | Tensile strength (N/m) | |
|---|---|---|---|---|
| | Armchair | Zigzag | Armchair | Zigzag |
| **N-GD-Hex** | 0.15 | 0.25 | 9.3 | 7.4 |
| **N-GD-Rec** | 0.75 | 0.24 | 13.3 | 26.2 |
| **N-GY-Hex** | 0.26 | 0.23 | 10.7 | 9.3 |
| **N-GY-Rec** | 0.6 | 0.21 | 13.5 | 27.6 |
| **B-GD-Hex** | 0.87 | 0.59 | 14.9 | 9.7 |
| **B-GD-Rec** | 0.48 | 0.27 | 13.9 | 13.7 |
| **B-GY-Hex** | 0.23 | 0.33 | 9.3 | 14.2 |
| **B-GY-Rec** | 0.27 | 0.27 | 10.2 | 18.5 |
| **P-GD-Hex** | 0.32 | 0.22 | 15.3 | 5.3 |
| **P-GD-Rec** | 0.79 | 0.21 | 24.0 | 16.1 |
| **P-GY-Hex** | 0.32 | 0.22 | 14.2 | 6.6 |
| **P-GY-Rec** | 0.66 | 0.18 | 23.2 | 15.7 |
| **Al-GD-Hex** | 0.38 | 0.32 | 8.4 | 6.5 |
| **Al-GD-Rec** | 0.37 | 0.27 | 8.5 | 9.1 |
| **Al-GY-Hex** | 0.36 | 035 | 8.8 | 8.8 |
| **Al-GY-Rec** | 0.31 | 0.27 | 9.1 | 12.0 |
| **As-GD-Hex** | 0.31 | 0.21 | 8.2 | 3.9 |
| **As-GD-Rec** | 0.75 | 0.21 | 23.9 | 11.6 |
| **As-GY-Hex** | 0.31 | 0.21 | 10.4 | 5.1 |
| **As-GY-Rec** | 0.53 | 0.16 | 18.7 | 10.0 |
| **Ga-GD-Hex** | 0.36 | 0.32 | 8.2 | 6.1 |
| **Ga-GD-Rec** | 0.37 | 0.26 | 8.3 | 8.6 |
| **Ga-GY-Hex** | 0.35 | 0.33 | 8.5 | 8.1 |
| **Ga-GY-Rec** | 0.29 | 0.26 | 8.9 | 11.4 |

Remarkable mechanical properties of predicated nanosheets are not adequate to ensure their thermal stability. In addition, for most of the applications the components should be able to stay intact at high temperatures in order to avoid failure upon sudden or continuous temperature rises. To explore the thermal stability of predicted nanosheets, we conducted the AIMD simulations at the high temperatures of 1000 K and 2000 K for 20 ps. According to our results (shown in Fig. S1), almost all the constructed nanomembranes could stay intact at the high temperature of 2000 K. Nonetheless, As-graphyne/graphdiyne nanosheets with the graphene like lattices were found to disintegrate at 2000 K, however, they could stay fully intact at 1000 K. The AIMD results confirm the remarkable thermal stability of predicted graphyne/graphdiyne nanosheets.

Table 3, Estimated energy band-gap (eV), first adsorption peak of Im ε (eV) and static dielectric constant of single-layer graphyne/graphdiyne lattices.

| Structure | Band-gap | | First adsorption peak of Im ε (eV) | | Static dielectric constant | |
|---|---|---|---|---|---|---|
| | PBE | HSE06 | E\|\|x | E\|\|y | E\|\|x | E\|\|y |
| Al-GD-Hex | 2.03 | 2.94 | 2.70 | 2.70 | 1.56 | 1.56 |
| Al-GY-Hex | 2.61 | 3.73 | 3.37 | 3.37 | 1.46 | 1.46 |
| Al-GD-Rec | 0.96 | 1.55 | 1.33 | 2.02 | 2.32 | 2.13 |
| Al-GY-Rec | 1.19 | 1.89 | 1.76 | 2.79 | 2.15 | 1.88 |
| As-GD-Hex | 1.12 | 1.55 | 1.25 | 1.25 | 3.13 | 3.13 |
| As-GY-Hex | 1.19 | 1.59 | 1.29 | 1.29 | 3.50 | 3.50 |
| As-GD-Rec | 0.47 | 0.75 | 0.51 | 0.97 | 5.15 | 7.25 |
| As-GY-Rec | 0.30 | 0.57 | 0.32 | 0.52 | 10.13 | 8.67 |
| B-GD-Hex | 0.47 | 1.33 | 1.40 | 1.40 | 2.42 | 2.42 |
| B-GY-Hex | 0.39 | 1.43 | 1.63 | 1.63 | 2.18 | 2.18 |
| B-GD-Rec | 0.17 | 0.75 | 0.71 | 1.13 | 4.49 | 3.43 |
| B-GY-Rec | 0.01 | 0.58 | 0.75 | 1.44 | 5.23 | 3.42 |
| Ga-GD-Hex | 1.96 | 2.93 | 2.56 | 2.56 | 1.61 | 1.61 |
| Ga-GY-Hex | 2.57 | 3.66 | 3.26 | 3.26 | 1.48 | 1.48 |
| Ga-GD-Rec | 0.86 | 1.45 | 1.29 | 1.94 | 2.34 | 2.18 |
| Ga-GY-Rec | 1.06 | 1.79 | 1.64 | 2.75 | 2.18 | 1.90 |
| P-GD-Hex | 1.37 | 1.90 | 1.48 | 1.48 | 2.67 | 2.67 |
| P-GY-Hex | 1.48 | 2.01 | 1.56 | 1.56 | 2.85 | 2.85 |
| P-GD-Rec | 0.35 | 0.76 | 0.44 | 1.13 | 6.16 | 6.65 |
| P-GY-Rec | 0.23 | 0.49 | 0.24 | 1.01 | 18.10 | 7.12 |
| N-GD-Hex | 0.79 | 1.27 | 2.53 | 2.53 | 1.66 | 1.66 |
| N-GY-Hex | 2.52 | 4.00 | 3.11 | 3.11 | 1.65 | 1.65 |
| N-GD-Rec | 1.63 | 2.23 | 0.55 | 1.46 | 3.21 | 3.99 |
| N-GY-Rec | 0.39 | 0.89 | 0.40 | 1.20 | 6.55 | 5.71 |

In order to probe the electrical characteristics of the predicted graphdiyne/graphyne nanosheets, the band structures along the high symmetry directions were calculated and the acquired results are depicted in Figs. 3 and 4. Obtained results shown in Fig. 3 indicate that both valence band maximum (VBM) and the conduction band minimum (CBM) of all B-, Al-, and Ga-graphdiyne/graphyne monolayers with the hexagonal lattices occur at the *G*-point, showing direct band-gaps. The VBM for these structures with the rectangular phases coincide at *S-Y* direction while the CBM for the graphdiyne structures locate at *X*-point and for the graphyne structures occur at *G*-point, showing indirect band-gaps.

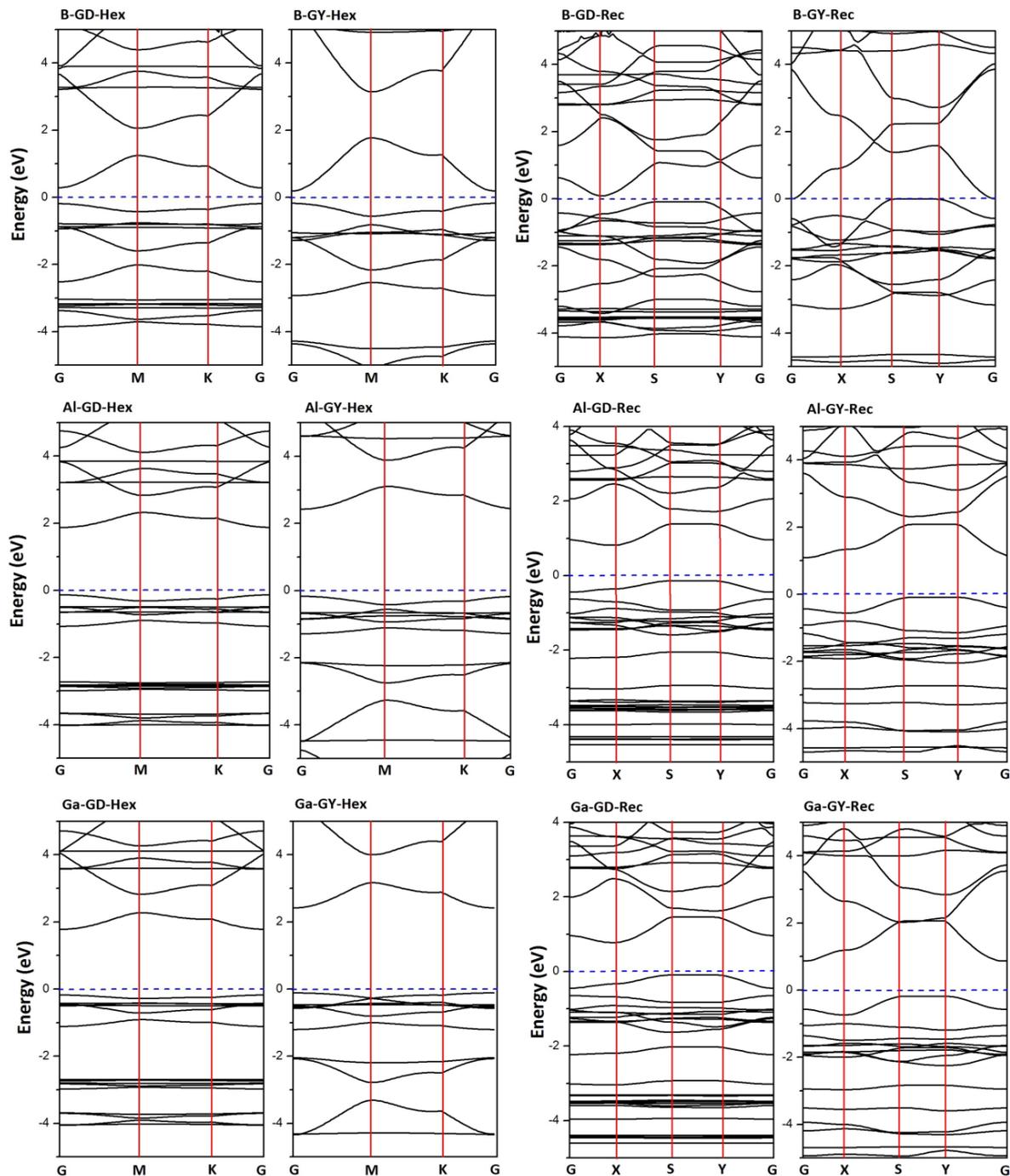

**Fig. 3**, Band structures of single-layers B-, Al- and Ga-graphdiyne/graphyne monolayers with hexagonal and rectangular atomic lattices predicted by the PBE functional. The Fermi energy is aligned to zero.

Fig. 4 illustrates the band structures for the N-, P- and As-graphyne/graphdiyne monolayers with the hexagonal and rectangular phases. These aforementioned monolayers with both hexagonal and rectangular atomic lattices are found to be direct band-gap semiconductors, since the VBM and CBM coincide at the *G*-point.

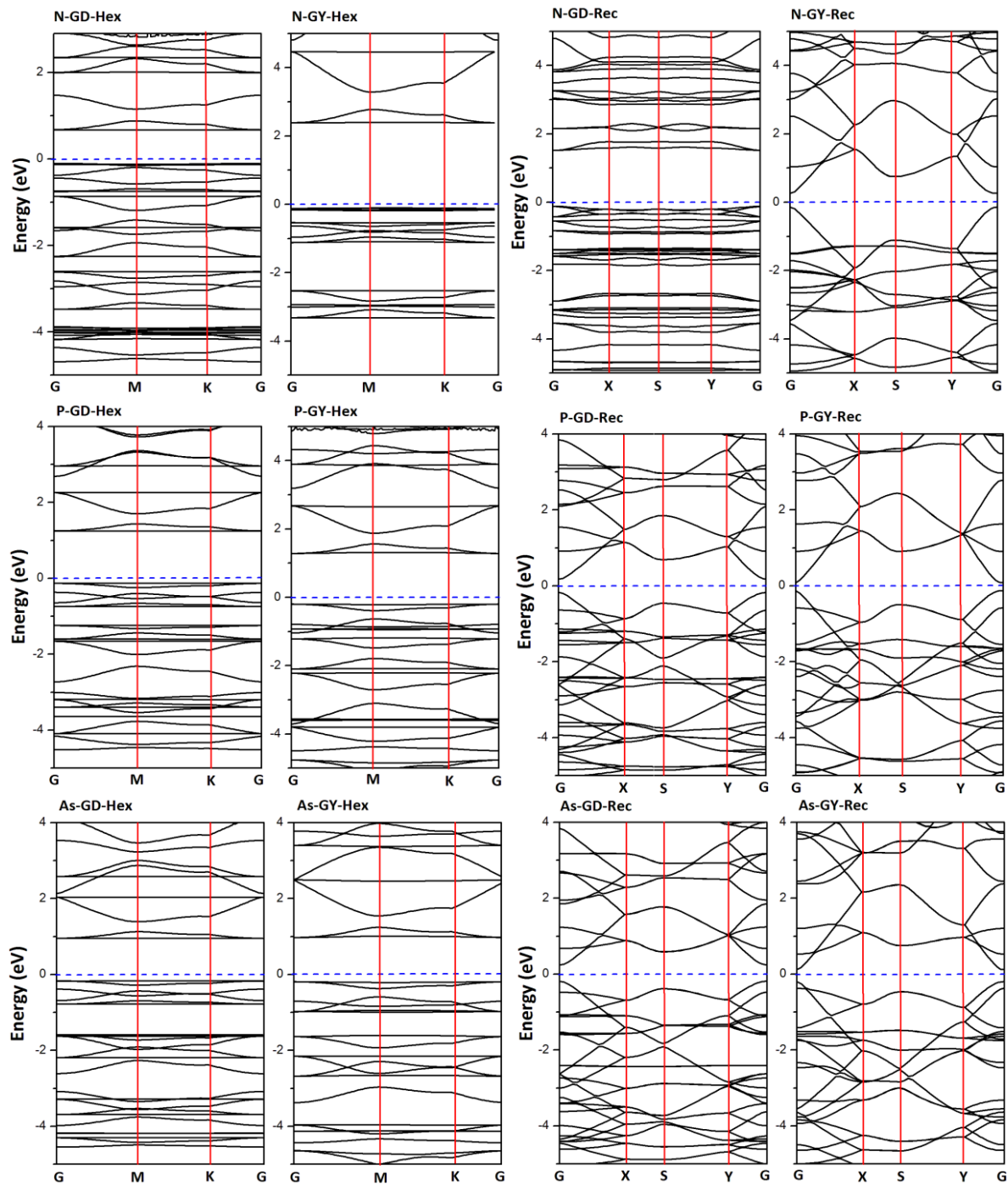

**Fig. 4**, Band structures of single-layers N-, P- and As-graphdiyne/graphyne monolayers with hexagonal and rectangular atomic lattices predicted by the PBE functional. The Fermi energy is aligned to zero.

As the PBE/GGA functional is well-known to underestimate the band-gap values of semiconductors, the electronic density of states were also computed using the HSE06 hybrid functional and the obtained results are depicted in Fig. S2. Table 3 summarizes the acquired results for the band-gap values predicted within the PBE/GGA and HSE06 approaches. The semiconducting band-gap values of N-GD, P-GD, As-GD, N-GY, P-GY and As-GY monolayers with the hexagonal lattices are 1.27 eV, 1.90 eV, 1.55 eV, 4.00 eV, 2.01 eV and 1.59 eV while the corresponding band-gaps for the same structures with the rectangular lattices are 2.23 eV, 0.76 eV, 0.75 eV, 0.89 eV, 0.49 eV and 0.57 eV, respectively. Worthy to note that the optimum band-gap of a photovoltaic device should be in a range between 1.1 eV and 1.4 eV (for a simple p–n junction), and for the water splitting application the band-gap should be ideally between 1.8 eV and 2.2 eV [68,69]. Hence B-GD-Hex and N-GD-Hex nanosheets might be suitable for the photovoltaic devices and B-GY-Hex, Al-GY-Rec, P-GD-Hex and P-GY-Hex may serve as promising candidates for the water splitting applications.

Since our electronic calculation results indicate the semiconducting electronic responses with band-gap values in a wide range, from 0.49 eV up to 4.00 eV, investigation of optical properties of these nanomaterials can be appealing for their practical applications in optoelectronics. Because of the depolarization effect in planar geometry for the light polarization perpendicular to the plane (out-of-plane) [70], we only focus on the optical absorption spectrum for the light polarization parallel to the plane (in-plane). The predicted monolayers with hexagonal atomic lattices have isotropic structures while those with rectangular phases yield highly anisotropic structures along the x and y directions, hence the electrons and photons in these systems illustrate highly anisotropic nature. Therefore, the imaginary and real parts of the dielectric function for the parallel polarized directions (E||x and E||y) have been calculated and the obtained results are illustrated in Fig. 5. The first adsorption peak of Im($\varepsilon$) for the all studied monolayers are summarized in Table 3. Our results show that for Al-GD-Hex, Al-GY-Hex, Al-GD-Rec, Al-GY-Rec, B-GY-Hex, Ga-GD-Hex, Ga-GY-Hex, Ga-GD-Rec, Ga-GY-Rec, P-GY-Hex, N-GD-Hex and N-GY-Hex monolayers the first absorption peaks of Im ($\varepsilon$) along the in-plane directions are between 1.50 eV and 3.50 eV which are in the visible range of light. These results indicate that these aforementioned nanosheets can absorb the visible light, highly desirable for the practical applications in the optoelectronic devices operating in the visible spectral range. The first absorption peaks of Im($\varepsilon$) for the rest of predicted monolayers are located in the energy range smaller than 1.50 eV, which are within the IR and near-IR (NIR) range of light. Table 3 also include the static dielectric constant, Re $\varepsilon(0)$, for the constructed monolayers. It is clear that the monolayers with the hexagonal phases have the same value of Re $\varepsilon$ (0) along the x and y directions, confirming that the optical response is isotropic along the in-plane directions. On the other side for the lattices with the rectangular lattices, they illustrate different values of Re ε(0) along the x and y directions. Highly anisotropic optical properties along the in-plane (E||x and E||y) directions can be promising for the design of novel electronic and optical nanodevices that exploit the in-plane anisotropic optical properties, such as the polarization-sensitive photodetectors.

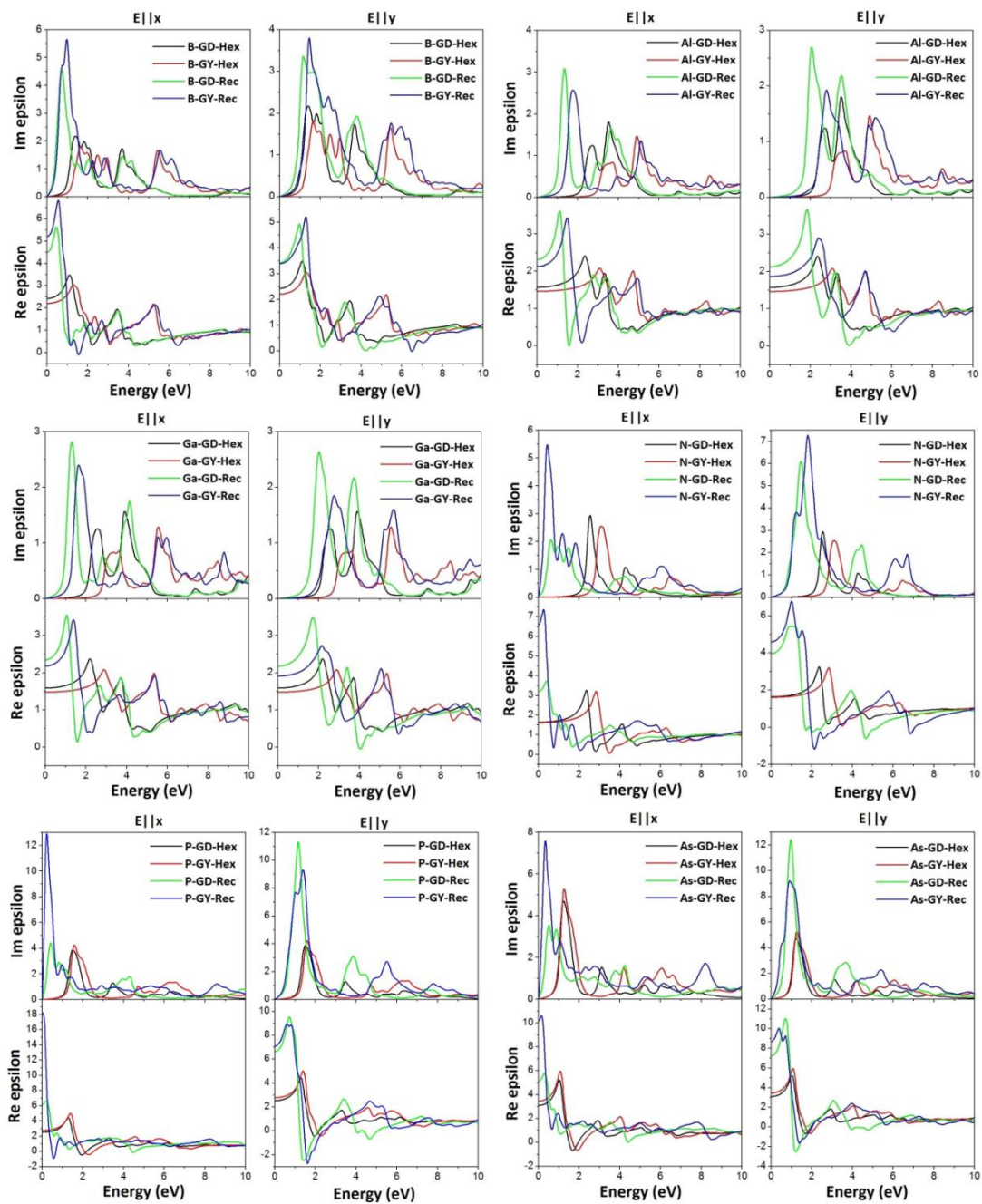

**Fig. 5**, Imaginary and real parts of the dielectric function of single-layer N-, B-, P-, Al-, As-, Ga-graphdiyne (GD) or -graphyne (GY) monolayers for the light polarizations along the E||x and E||y planar polarization directions, calculated using the RPA+PBE approach.

## 4. Concluding remarks

In this work, motivated by the latest experimental advances with respect to the synthesise of full carbon-, B- and N-graphdiyne nanomembranes, we predicted X-graphdiyne (X-GD) and X–graphyne (X-GY)(X= B, Al, Ga, N, P, As) monolayers with two different hexagonal and rectangular atomic lattices. We then explored the thermal stability, mechanical properties and electronic/optical characteristics of these novel nanosheets using the first-principles density functional theory calculations. It was confirmed that the predicted porous nanosheets can exhibit remarkably high tensile strengths or stretchability. Ab-initio molecular dynamics simulations confirm the outstanding thermal stability of all predicted monolayers. Our analysis also reveals highly attractive electronic and optical properties of predicted graphdiyne/graphyne nanosheets. It was found that predicted porous nanosheets exhibit semiconducting electronic characters with direct/indirect band-gaps, ranging from 0.49 eV up to 4.00 eV according to the HSE06 method estimations. The linear photon energy-dependent dielectric functions of the considered nanomaterials were investigated within the RPA+PBE approach. The first absorption peak of imaginary part of the dielectric function reveals that these novel nanomembranes can absorb the visible, IR and NIR light, illustrating their application prospect for the use in optoelectronics and nanoelectronics. On the other hand, highly anisotropic in-plane optical responses of predicted nanosheets with the rectangular atomic lattices suggest them as promising candidates for the polarization-sensitive photodetectors. The acquired results on the basis of first-principles simulations reveal the outstanding properties of predicted graphdiyne/graphyne porous lattices, and recommend them as highly promising candidates for the design of stretchable nanoelectronics and energy storage systems.

**Acknowledgment**

B. M. and T. R. acknowledge the financial support by European Research Council for COMBAT project (Grant number 615132). B. M. and X. Z. particularly appreciate the funding by the Deutsche Forschungsgemeinschaft (DFG, German Research Foundation) under Germany's Excellence Strategy within the Cluster of Excellence PhoenixD (EXC 2122, Project ID 390833453). T. H. is indebted to the resources at NCI National Facility systems at the Australian National University.

**Data availability**

The energy minimized atomic lattices are available to download from:

**References**

[1] K.S. Novoselov, A.K. Geim, S. V Morozov, D. Jiang, Y. Zhang, S. V Dubonos, et al., Electric field effect in atomically thin carbon films., Science. 306 (2004) 666–9. doi:10.1126/science.1102896.

[2] A.K. Geim, K.S. Novoselov, The rise of graphene, Nat. Mater. 6 (2007) 183–191. doi:10.1038/nmat1849.

[3] C. Lee, X. Wei, J.W. Kysar, J. Hone, Measurement of the Elastic Properties and Intrinsic Strength of Monolayer Graphene, Science (80-. ). 321 (2008) 385–388. doi:10.1126/science.1157996.

[4] A.A. Balandin, S. Ghosh, W. Bao, I. Calizo, D. Teweldebrhan, F. Miao, et al., Superior thermal conductivity of single-layer graphene, Nano Lett. 8 (2008) 902–907. doi:10.1021/nl0731872.

[5] A.A. Balandin, Thermal properties of graphene and nanostructured carbon materials, Nat. Mater. 10 (2011) 569–581. doi:10.1038/nmat3064.

[6] C. Berger, Z. Song, T. Li, X. Li, A.Y. Ogbazghi, R. Feng, et al., Ultrathin Epitaxial Graphite:  2D Electron Gas Properties and a Route toward Graphene-based Nanoelectronics, J. Phys. Chem. B. 108 (2004) 19912–19916. doi:doi:10.1021/jp040650f.

[7] M. Liu, X. Yin, E. Ulin-Avila, B. Geng, T. Zentgraf, L. Ju, et al., A graphene-based broadband optical modulator, Nature. 474 (2011) 64–67. doi:10.1038/nature10067.

[8] F. Withers, M. Dubois, A.K. Savchenko, Electron properties of fluorinated single-layer graphene transistors, Phys. Rev. B - Condens. Matter Mater. Phys. 82 (2010). doi:10.1103/PhysRevB.82.073403.

[9] B. Liu, K. Zhou, Recent progress on graphene-analogous 2D nanomaterials: Properties, modeling and applications, Prog. Mater. Sci. 100 (2019) 99–169. doi:10.1016/J.PMATSCI.2018.09.004.

[10] B. Radisavljevic, A. Radenovic, J. Brivio, V. Giacometti, A. Kis, Single-layer MoS2 transistors, Nat. Nanotechnol. 6 (2011) 147–50. doi:10.1038/nnano.2010.279.

[11] Q.H. Wang, K. Kalantar-Zadeh, A. Kis, J.N. Coleman, M.S. Strano, Electronics and optoelectronics of two-dimensional transition metal dichalcogenides, Nat. Nanotechnol. 7 (2012) 699–712. doi:10.1038/nnano.2012.193.

[12] H. Farahani, A. Rajabpour, M. Khanaki, A. Reyhani, Interfacial thermal resistance between few-layer MoS2 and silica substrates: A molecular dynamics study, Comput. Mater. Sci. 142 (2018) 1–6. doi:https://doi.org/10.1016/j.commatsci.2017.09.052.

[13] S. Das, M. Demarteau, A. Roelofs, Ambipolar phosphorene field effect transistor, ACS Nano. 8 (2014) 11730–11738. doi:10.1021/nn505868h.

[14] L. Li, Y. Yu, G.J. Ye, Q. Ge, X. Ou, H. Wu, et al., Black phosphorus field-effect transistors, Nat. Nanotechnol. 9 (2014) 372–377. doi:10.1038/nnano.2014.35.

[15] G. Algara-Siller, N. Severin, S.Y. Chong, T. Björkman, R.G. Palgrave, A. Laybourn, et al., Triazine-based graphitic carbon nitride: A two-dimensional semiconductor, Angew. Chemie - Int. Ed. 53 (2014) 7450–7455. doi:10.1002/anie.201402191.

[16] L.-B. Shi, Y.-Y. Zhang, X.-M. Xiu, H.-K. Dong, Structural characteristics and strain behaviors of two-dimensional C3N : First principles calculations, Carbon N. Y. 134 (2018) 103–111. doi:https://doi.org/10.1016/j.carbon.2018.03.076.

[17] L. Bin Shi, S. Cao, J. Zhang, X.M. Xiu, H.K. Dong, Mechanical behaviors and electronic characteristics on two-dimensional C2N3and C2N3H: First principles calculations, Phys. E Low-Dimensional Syst. Nanostructures. 103 (2018) 252–263. doi:10.1016/j.physe.2018.06.014.

[18] H.L. Lee, Z. Sofer, V. Mazánek, J. Luxa, C.K. Chua, M. Pumera, Graphitic carbon nitride: Effects of various precursors on the structural, morphological and electrochemical sensing properties, Appl. Mater. Today. 8 (2017) 150–162. doi:10.1016/j.apmt.2016.09.019.

[19] M. Makaremi, B. Mortazavi, C.V. Singh, 2D Hydrogenated graphene-like borophene as a high capacity anode material for improved Li/Na ion batteries: A first principles study, Mater. Today Energy. 8 (2018) 22–28. doi:10.1016/j.mtener.2018.02.003.

[20] A. Rajabpour, S. Bazrafshan, S. Volz, Carbon-nitride 2D nanostructures: Thermal conductivity and interfacial thermal conductance with the silica substrate, Phys. Chem. Chem. Phys. (2019). doi:10.1039/C8CP06992A.

[21] R.H. Baughman, H. Eckhardt, M. Kertesz, Structure-property predictions for new planar forms of carbon: Layered phases containing sp$^{2}$ and sp atoms, J. Chem. Phys. 87 (1987) 6687. doi:10.1063/1.453405.

[22] G. Li, Y. Li, H. Liu, Y. Guo, Y. Li, D. Zhu, Architecture of graphdiyne nanoscale films, Chem. Commun. 46 (2010) 3256–3258. doi:10.1039/B922733D.

[23] H. Shang, Z. Zuo, L. Li, F. Wang, H. Liu, Y. Li, et al., Ultrathin Graphdiyne Nanosheets Grown In Situ on Copper Nanowires and Their Performance as Lithium-Ion Battery Anodes, Angew. Chemie Int. Ed. 57 (2017) 774–778. doi:10.1002/anie.201711366.

[24] H. Shang, Z. Zuo, L. Yu, F. Wang, F. He, Y. Li, Low-Temperature Growth of All-Carbon Graphdiyne on a Silicon Anode for High-Performance Lithium-Ion Batteries, Adv. Mater. 30 (2018) 1801459. doi:10.1002/adma.201801459.

[25] B. Mortazavi, M. Shahrokhi, X. Zhuang, T. Rabczuk, Boron-graphdiyne: A superstretchable semiconductor with low thermal conductivity and ultrahigh capacity for Li, Na and Ca ion

storage, J. Mater. Chem. A. 6 (2018) 11022–11036. doi:10.1039/c8ta02627k.

[26] B. Mortazavi, M. Shahrokhi, M.E. Madjet, M. Makaremi, S. Ahzi, T. Rabczuk, N-, P-, As-triphenylene-graphdiyne: Strong and stable 2D semiconductors with outstanding capacities as anodes for Li-ion batteries, Carbon N. Y. 141 (2019) 291–303. doi:https://doi.org/10.1016/j.carbon.2018.09.070.

[27] M. Salavati, T. Rabczuk, Application of highly stretchable and conductive two-dimensional 1T VS2 and VSe2 as anode materials for Li-, Na- and Ca-ion storage, Comput. Mater. Sci. 160 (2019) 360–367. doi:https://doi.org/10.1016/j.commatsci.2019.01.018.

[28] L.D. Pan, L.Z. Zhang, B.Q. Song, S.X. Du, H.J. Gao, Graphyne- and graphdiyne-based nanoribbons: Density functional theory calculations of electronic structures, Appl. Phys. Lett. 98 (2011). doi:10.1063/1.3583507.

[29] M. Becton, L. Zhang, X. Wang, Mechanics of graphyne crumpling, Phys. Chem. Chem. Phys. (2014). doi:10.1039/c4cp02400a.

[30] C. Lenear, M. Becton, X. Wang, Computational analysis of hydrogenated graphyne folding, Chem. Phys. Lett. (2016). doi:10.1016/j.cplett.2016.01.025.

[31] W. Zhang, J.K. Huang, C.H. Chen, Y.H. Chang, Y.J. Cheng, L.J. Li, High-gain phototransistors based on a CVD MoS2 monolayer, Adv. Mater. 25 (2013) 3456–3461. doi:10.1002/adma.201301244.

[32] Z. Xu, X. Lv, J. Li, J. Chen, Q. Liu, A promising anode material for sodium-ion battery with high capacity and high diffusion ability: Graphyne and graphdiyne, RSC Adv. 6 (2016) 25594–25600. doi:10.1039/c6ra01870j.

[33] T. Hussain, M. Hankel, D.J. Searles, Graphenylene Monolayers Doped with Alkali or Alkaline Earth Metals: Promising Materials for Clean Energy Storage, J. Phys. Chem. C. 121 (2017) 14393–14400. doi:10.1021/acs.jpcc.7b02191.

[34] M. Bartolomei, E. Carmona-Novillo, G. Giorgi, First principles investigation of hydrogen physical adsorption on graphynes' layers, Carbon N. Y. 95 (2015) 1076–1081. doi:10.1016/j.carbon.2015.08.118.

[35] P.A.S. Autreto, J.M. De Sousa, D.S. Galvao, Site-dependent hydrogenation on graphdiyne, Carbon N. Y. 77 (2014) 829–834. doi:10.1016/j.carbon.2014.05.088.

[36] Z.Z. Lin, Graphdiyne as a promising substrate for stabilizing Pt nanoparticle catalyst, Carbon N. Y. 86 (2015) 301–309. doi:10.1016/j.carbon.2015.02.014.

[37] L. Sun, P.H. Jiang, H.J. Liu, D.D. Fan, J.H. Liang, J. Wei, et al., Graphdiyne: A two-dimensional thermoelectric material with high figure of merit, Carbon N. Y. 90 (2015) 255–259. doi:10.1016/j.carbon.2015.04.037.

[38] X.M. Wang, S.S. Lu, Thermoelectric transport in graphyne nanotubes, J. Phys. Chem. C. 117 (2013) 19740–19745. doi:10.1021/jp406536e.

[39] R. Matsuoka, R. Sakamoto, K. Hoshiko, S. Sasaki, H. Masunaga, K. Nagashio, et al., Crystalline Graphdiyne Nanosheets Produced at a Gas/Liquid or Liquid/Liquid Interface, J. Am. Chem. Soc. 139 (2017) 3145–3152. doi:10.1021/jacs.6b12776.

[40] X. Kan, Y. Ban, C. Wu, Q. Pan, H. Liu, J. Song, et al., Interfacial Synthesis of Conjugated Two-Dimensional N-Graphdiyne, ACS Appl. Mater. Interfaces. 10 (2018) 53–58. doi:10.1021/acsami.7b17326.

[41] N. Wang, X. Li, Z. Tu, F. Zhao, J. He, Z. Guan, et al., Synthesis, Electronic Structure of Boron-Graphdiyne with an sp-Hybridized Carbon Skeleton and Its Application in Sodium Storage, Angew. Chemie. (2018). doi:10.1002/ange.201801897.

[42] B. Mortazavi, M. Makaremi, M. Shahrokhi, Z. Fan, T. Rabczuk, N-graphdiyne two-dimensional nanomaterials: Semiconductors with low thermal conductivity and high stretchability, Carbon N. Y. 137 (2018) 57–67. doi:10.1016/j.carbon.2018.04.090.

[43] M. Makaremi, B. Mortazavi, T. Rabczuk, G.A. Ozin, C.V. Singh, Theoretical Investigation: 2D N-Graphdiyne Nanosheets as Promising Anode Materials for Li/Na Rechargeable Storage Devices, ACS Appl. Nano Mater. 2 (2019) 127–135. doi:10.1021/acsanm.8b01751.

[44] B. Mortazavi, M. Shahrokhi, M. Makaremi, T. Rabczuk, Anisotropic mechanical and optical response and negative Poisson's ratio in $Mo_2C$ nanomembranes revealed by first-principles simulations, Nanotechnology. 28 (2017). doi:10.1088/1361-6528/aa5c29.

[45] A.R. Oganov, C.W. Glass, Crystal structure prediction using ab initio evolutionary techniques: principles and applications., J. Chem. Phys. 124 (2006) 244704. doi:10.1063/1.2210932.

[46] C.W. Glass, A.R. Oganov, N. Hansen, USPEX-Evolutionary crystal structure prediction, Comput. Phys. Commun. 175 (2006) 713–720. doi:10.1016/j.cpc.2006.07.020.

[47] L.-B. Shi, S. Cao, M. Yang, Strain behavior and Carrier mobility for novel two-dimensional semiconductor of GeP: First principles calculations, Phys. E Low-Dimensional Syst. Nanostructures. 107 (2019) 124–130. doi:10.1016/J.PHYSE.2018.11.024.

[48] G. Kresse, J. Furthm??ller, Efficiency of ab-initio total energy calculations for metals and semiconductors using a plane-wave basis set, Comput. Mater. Sci. 6 (1996) 15–50. doi:10.1016/0927-0256(96)00008-0.

[49] G. Kresse, J. Furthmüller, Efficient iterative schemes for ab initio total-energy calculations using a plane-wave basis set, Phys. Rev. B. 54 (1996) 11169–11186. doi:10.1103/PhysRevB.54.11169.

[50] G. Kresse, From ultrasoft pseudopotentials to the projector augmented-wave method, Phys.

Rev. B. 59 (1999) 1758–1775. doi:10.1103/PhysRevB.59.1758.

[51] J. Perdew, K. Burke, M. Ernzerhof, Generalized Gradient Approximation Made Simple., Phys. Rev. Lett. 77 (1996) 3865–3868. doi:10.1103/PhysRevLett.77.3865.

[52] P.E. Blöchl, Projector augmented-wave method, Phys. Rev. B. 50 (1994) 17953–17979. doi:10.1103/PhysRevB.50.17953.

[53] K. Momma, F. Izumi, VESTA 3 for three-dimensional visualization of crystal, volumetric and morphology data, J. Appl. Crystallogr. 44 (2011) 1272–1276. doi:10.1107/S0021889811038970.

[54] B. Mortazavi, M. Shahrokhi, M. Makaremi, G. Cuniberti, T. Rabczuk, First-principles investigation of Ag-, Co-, Cr-, Cu-, Fe-, Mn-, Ni-, Pd- and Rh-hexaaminobenzene 2D metal-organic frameworks, Mater. Today Energy. 10 (2018) 336–342. doi:10.1016/J.MTENER.2018.10.007.

[55] A.V.K. and O.A.V. and A.F.I. and G.E. Scuseria, Influence of the exchange screening parameter on the performance of screened hybrid functionals, J. Chem. Phys. 125 (2006) 224106. doi:10.1063/1.2404663.

[56] H. Monkhorst, J. Pack, Special points for Brillouin zone integrations, Phys. Rev. B. 13 (1976) 5188–5192. doi:10.1103/PhysRevB.13.5188.

[57] M. Shahrokhi, Quasi-particle energies and optical excitations of ZnS monolayer honeycomb structure, Appl. Surf. Sci. 390 (2016) 377–384. doi:http://dx.doi.org/10.1016/j.apsusc.2016.08.055.

[58] M. Shahrokhi, C. Leonard, Quasi-particle energies and optical excitations of wurtzite BeO and its nanosheet, J. Alloys Compd. 682 (2016) 254–262. doi:http://dx.doi.org/10.1016/j.jallcom.2016.04.288.

[59] M. Shahrokhi, Can fluorine and chlorine functionalization stabilize the graphene like borophene?, Comput. Mater. Sci. 156 (2019) 56–66. doi:10.1016/j.commatsci.2018.09.045.

[60] J. Mahmood, E.K. Lee, M. Jung, D. Shin, I.-Y. Jeon, S.-M. Jung, et al., Nitrogenated holey two-dimensional structures, Nat. Commun. 6 (2015) 6486. doi:10.1038/ncomms7486.

[61] B. Silvi, A. Savin, Classification of Chemical-Bonds Based on Topological Analysis of Electron Localization Functions, Nature. 371 (1994) 683–686. doi:10.1038/371683a0.

[62] A.H.N. Shirazi, Molecular dynamics investigation of mechanical properties of single-layer phagraphene, Front. Struct. Civ. Eng. (2018). doi:10.1007/s11709-018-0492-4.

[63] A.H.N. Shirazi, R. Abadi, M. Izadifar, N. Alajlan, T. Rabczuk, Mechanical responses of pristine and defective C3N nanosheets studied by molecular dynamics simulations, Comput. Mater. Sci. 147 (2018) 316–321. doi:10.1016/j.commatsci.2018.01.058.

[64] S. Sadeghzadeh, Effects of vacancies and divacancies on the failure of C3N nanosheets, Diam. Relat. Mater. 89 (2018) 257–265. doi:10.1016/J.DIAMOND.2018.09.018.

[65] M. Salavati, Electronic and mechanical responses of two-dimensional HfS2, HfSe2, ZrS2, and ZrSe2 from first-principles, Front. Struct. Civ. Eng. (2018) 1–9. doi:10.1007/s11709-018-0491-5.

[66] F. Liu, P. Ming, J. Li, Ab initio calculation of ideal strength and phonon instability of graphene under tension, Phys. Rev. B - Condens. Matter Mater. Phys. 76 (2007). doi:10.1103/PhysRevB.76.064120.

[67] B. Mortazavi, O. Rahaman, M. Makaremi, A. Dianat, G. Cuniberti, T. Rabczuk, First-principles investigation of mechanical properties of silicene, germanene and stanene, Phys. E Low-Dimensional Syst. Nanostructures. 87 (2017) 228–232. doi:10.1016/j.physe.2016.10.047.

[68] Y. Moriya, T. Takata, K. Domen, Recent progress in the development of (oxy)nitride photocatalysts for water splitting under visible-light irradiation, Coord. Chem. Rev. (2013). doi:10.1016/j.ccr.2013.01.021.

[69] T. Le Bahers, M. R??rat, P. Sautet, Semiconductors used in photovoltaic and photocatalytic devices: Assessing fundamental properties from DFT, J. Phys. Chem. C. (2014). doi:10.1021/jp409724c.

[70] M. Shahrokhi, Quasi-particle energies and optical excitations of ZnS monolayer honeycomb structure, Appl. Surf. Sci. 390 (2016) 377–384. doi:10.1016/j.apsusc.2016.08.055.

# Supporting Information

# N-, B-, P-, Al-, As-, Ga-graphdiyne/graphyne lattices; First-principles investigation of mechanical, optical and electronic properties

*E-mail: bohayra.mortazavi@gmail.com

1. Atomic structures of constructed monolayers unit-cells in VASP POSCAR.

2- AIMD results for the thermal stability

3- HSE06 results for the total electronic density of states.

```
N-GD-Hex.
   1.00000000000000
     16.0375168077986103    0.0000000000000000    0.0000000000000000
      8.0187584038993034   13.8888969726305600    0.0000000000000000
      0.0000000000000000    0.0000000000000000   20.0000000000000000
   C    N
   18    6
Direct
  0.3804121860736274  0.3804121974542980  0.5000000000000000
  0.4316795762031731  0.4316795914129870  0.5000000000000000
  0.4757496464837061  0.4757496360660340  0.5000000000000000
  0.5242503541734678  0.5242503770415965  0.5000000000000000
  0.5683204244540008  0.5683204216946365  0.5000000000000000
  0.6195877851269032  0.6195878156533254  0.5000000000000000
  0.2391756153880706  0.3804121974542980  0.5000000000000000
  0.3804121899027848  0.2391756250017565  0.5000000000000000
  0.6195878141557571  0.7608243813031450  0.5000000000000000
  0.7608243499211298  0.6195878156533254  0.5000000000000000
  0.5683204201970540  0.8633591692205371  0.5000000000000000
  0.5242503755440140  0.9514992585266029  0.5000000000000000
  0.4316795872628418  0.1366408302816495  0.5000000000000000
  0.4757496319158818  0.0485007409755767  0.5000000000000000
  0.1366429261230877  0.4316795846102650  0.5000000000000000
  0.0485049320680702  0.4757496292633121  0.5000000000000000
  0.8633570459888276  0.5683204148919145  0.5000000000000000
  0.9514950400438522  0.5242503702388746  0.5000000000000000
  0.2833474513407666  0.4333051021258214  0.5000000000000000
  0.4333050907451508  0.2833474588922726  0.5000000000000000
  0.2833474492785868  0.2833474588922726  0.5000000000000000
  0.5666948838567194  0.7166525474126288  0.5000000000000000
  0.7166525493164073  0.5666949109818020  0.5000000000000000
  0.7166525194319675  0.7166525474126288  0.5000000000000000
```

```
N-GY-Hex.
    11.5797873518352397     0.0000000000000000     0.0000000000000000
     5.7898936759176269    10.0283900171084106     0.0000000000000000
     0.0000000000000000     0.0000000000000000    20.0000000000000000
   C    N
   12    6
Direct
  0.3975461511808547  0.3987924258657713  0.5000000000000000
  0.4685431739356289  0.4702748313915350  0.5000000000000000
  0.6006470415036986  0.6018240656336928  0.5000000000000000
  0.5291969403822065  0.5308089384523669  0.5000000000000000
  0.2024805047568350  0.3981648927816792  0.5000000000000000
  0.3981417432226024  0.2030826264798407  0.5000000000000000
  0.6012319802712014  0.7969027701085238  0.5000000000000000
  0.7963442103420633  0.6011940784531404  0.5000000000000000
  0.5302420982084755  0.9394030545548517  0.5000000000000000
  0.4696462107871966  0.0605897311760231  0.5000000000000000
  0.0600113572813541  0.4691801693066679  0.5000000000000000
  0.9388256734944491  0.5296969456450924  0.5000000000000000
  0.2632619341083853  0.4716467956766834  0.5000000000000000
  0.4709938337485937  0.2645160471613082  0.5000000000000000
  0.2638737991634841  0.2638794235391089  0.5000000000000000
  0.5277589604892654  0.7361099311326029  0.5000000000000000
  0.7349096043595935  0.5283397175771967  0.5000000000000000
  0.7355193567640583  0.7354666850638765  0.5000000000000000
```

```
N-GD-Rec.
1.00000000000000
     9.6700000762999991    0.0000000000000000    0.0000000000000000
     0.0000000000000000   15.9700002669999996    0.0000000000000000
     0.0000000000000000    0.0000000000000000   20.0000000000000000
   C    N
   24    4
Direct
  0.1030000000000015  0.0890000019999988  0.5000000000000000
  0.1780000030000011  0.1640000049999983  0.5000000000000000
  0.2430000009999986  0.2300000039999972  0.5000000000000000
  0.3140000100000009  0.3030000030000011  0.5000000000000000
  0.3790000079999984  0.3689999879999988  0.5000000000000000
  0.4539999960000003  0.4440000060000031  0.5000000000000000
  0.9539999960000003  0.0890000019999988  0.5000000000000000
  0.8790000079999984  0.1640000049999983  0.5000000000000000
  0.8140000100000009  0.2300000039999972  0.5000000000000000
  0.7429999710000033  0.3030000030000011  0.5000000000000000
  0.6779999729999986  0.3689999879999988  0.5000000000000000
  0.6029999849999967  0.4440000060000031  0.5000000000000000
  0.4539999960000003  0.5889999870000011  0.5000000000000000
  0.3790000079999984  0.6639999750000030  0.5000000000000000
  0.3140000100000009  0.7300000190000020  0.5000000000000000
  0.2430000009999986  0.8029999729999986  0.5000000000000000
  0.1780000030000011  0.8690000180000013  0.5000000000000000
  0.1030000000000015  0.9440000060000031  0.5000000000000000
  0.6029999849999967  0.5889999870000011  0.5000000000000000
  0.6779999729999986  0.6639999750000030  0.5000000000000000
  0.7429999710000033  0.7300000190000020  0.5000000000000000
  0.8140000100000009  0.8029999729999986  0.5000000000000000
  0.8790000079999984  0.8690000180000013  0.5000000000000000
  0.9539999960000003  0.9440000060000031  0.5000000000000000
  0.1749999970000005  0.0160000010000019  0.5000000000000000
  0.8820000289999967  0.0160000010000019  0.5000000000000000
  0.3819999990000014  0.5159999729999996  0.5000000000000000
  0.6750000119999982  0.5159999729999996  0.5000000000000000
```

```
N-GY-Rec.
   1.00000000000000
     7.0518848930517954    0.0000000000000000    0.0000000000000000
     0.0000000000000000   11.5450439196077035    0.0000000000000000
     0.0000000000000000    0.0000000000000000   20.0000000000000000
   C    N
  16    4
Direct
  0.1391125793967873  0.1235654885620434  0.5000000000000000
  0.9347703032909891  0.1235659594382170  0.5000000000000000
  0.2429134365813752  0.2281861732100054  0.5000000000000000
  0.8309662910531941  0.2281855003073119  0.5000000000000000
  0.3309721022024235  0.3193156587373949  0.5000000000000000
  0.7429047944840406  0.3193139794365294  0.5000000000000000
  0.4347694865073137  0.4239371360481385  0.5000000000000000
  0.6391110308108736  0.4239370473975370  0.5000000000000000
  0.4347703249608159  0.6235647245859965  0.5000000000000000
  0.6391117186747195  0.6235650809339219  0.5000000000000000
  0.3309732943633961  0.7281871673125337  0.5000000000000000
  0.7429060847906044  0.7281879216405684  0.5000000000000000
  0.2429121988745067  0.8193157822493404  0.5000000000000000
  0.8309644509210514  0.8193175505700410  0.5000000000000000
  0.1391113567533750  0.9239359318246798  0.5000000000000000
  0.9347687722682707  0.9239367911429142  0.5000000000000000
  0.8367555659106998  0.0237512760830541  0.5000000000000000
  0.2371281563175671  0.0237508141460694  0.5000000000000000
  0.3367532281006902  0.5237515600530287  0.5000000000000000
  0.7371302038786638  0.5237504028537643  0.5000000000000000
```

```
B-GD-Hex
   1.00000000000000
     11.8467556014048103    0.0000000000000000    0.0000000000000000
      5.9233778007024069   10.2595913032290600    0.0000000000000000
      0.0000000000000000    0.0000000000000000   20.0000000000000000
    C  B
    12    2
Direct
   0.4070308672173653  0.1859346011964433  0.5000000006440217
   0.4672015211373619  0.0655895691018387  0.5000000029657059
   0.4070311110582381  0.4070304009524435  0.4999999994757971
   0.4672009784262485  0.4672011005079426  0.5000000001081091
   0.5327928226503715  0.5327934736097859  0.5000000019487061
   0.5929670091187731  0.5929683745784118  0.5000000023988420
   0.5929644137274650  0.8140626494910990  0.4999999968751183
   0.5327933412711943  0.9344053481882924  0.4999999985596588
   0.1859349549626330  0.4070305223908619  0.5000000039912749
   0.0655896108587868  0.4672031250811983  0.4999999994225561
   0.8140623904039757  0.5929635621580545  0.4999999967410673
   0.9344050922340088  0.5327923036735314  0.4999999982217460
   0.3333309009287149  0.3333311806515624  0.5000000025095019
   0.6666637191866585  0.6666639905117151  0.4999999961379018

   B-GY-Hex.
   1.00000000000000
      7.3993340266393819    0.0000000000000000    0.0000000000000000
      3.6996670133196909    6.4080112381514613    0.0000000000000000
      0.0000000000000000    0.0000000000000000   20.0000000000000000
    C  B
     6    2
Direct
   0.5479762480075649  0.9039220011690006  0.5000000793771591
   0.5479869980626120  0.5479852524329090  0.4999999399615902
   0.4519468433367706  0.4519467250748832  0.4999999686695276
   0.4520326477347396  0.0960011329332673  0.4999998768714633
   0.9039167495750249  0.5479814554799134  0.5000000725807894
   0.0960060592445231  0.4520289232261874  0.5000001081962822
   0.3333059913196195  0.3333054936785800  0.5000001525667130
   0.6666298759498659  0.6666305724160608  0.4999998017764611
```

```
    B-GD-Rec.
       15.1176650848092127    0.0000000000000000    0.0000000000000000
        0.0000000000000000   11.5612740620846104    0.0000000000000000
        0.0000000000000000    0.0000000000000000   20.0000000000000000
     C   B
      22    2
Direct
   0.4102543203183444   0.4972997911539707   0.4999943875085933
   0.4919637545246118   0.4970378879462274   0.4999940130849367
   0.0835927457478149   0.9973498433664432   0.5000095736013748
   0.5808820385932378   0.4972124811580088   0.4999599154717203
   0.6625666286417982   0.4979044381627347   0.4999522700811241
   0.8129567423530801   0.6107869238143309   0.5000175685562596
   0.8541040797492556   0.7031882062001102   0.5000189915735689
   0.8994224109006836   0.8029757052982518   0.5000033702213997
   0.9417467152681525   0.8940659224030654   0.4999986036754152
   0.8124522838408836   0.3849345856775699   0.5000014716709060
   0.8537998614889659   0.2926848723814501   0.5000045507176125
   0.8995672581886254   0.1932835705093723   0.5000075365614407
   0.9422300136765998   0.1024618047785353   0.5000108351598911
   0.9893746828101939   0.9979912298404514   0.5000120512789721
   0.2600157029430008   0.6101205945504731   0.4999991645283117
   0.2189736087037986   0.7025882507091126   0.5000124457462049
   0.1736745103528392   0.8023807938007934   0.5000142302960171
   0.1308458105754013   0.8930956778439594   0.5000139850056300
   0.2603278219382759   0.3842164513274966   0.5000008628663153
   0.2190016690507264   0.2919763568140610   0.4999997481206933
   0.1737068959249086   0.1922734938699264   0.4999995814372298
   0.1311633460361961   0.1013913330666938   0.5000066670304842
   0.3102668156786805   0.4972810803439316   0.4999980839611595
   0.7625361889960516   0.4980490806622413   0.4999897363013588

    B-GY-Rec.
      1.000000000000000
        9.9118335633718662    0.0000000000000000    0.0000000000000000
        0.0000000000000000    7.1456395344936281    0.0000000000000000
        0.0000000000000000    0.0000000000000000   20.0000000000000000
     C   B
      12    2
Direct
   0.4436304256107917   0.4977387292673399   0.5000007394621164
   0.5678475347842138   0.4979349552453982   0.5000018864417797
   0.7973140674481201   0.6820363819270483   0.5000027950606594
   0.0769959974217684   0.9972719175965992   0.5000018115432567
   0.8612945539651733   0.8292139281396160   0.5000029907859656
   0.7981220739329871   0.3155363053877949   0.4999904995003064
   0.8621625230843861   0.1683925052963104   0.4999942627648863
   0.9343430405121111   0.9984003996479487   0.5000002624103104
   0.1491270017164865   0.8273009720484268   0.5000015666116582
   0.2134506872780619   0.6803162790984132   0.5000017468919111
   0.2141025678304445   0.3136742697527382   0.5000013707553776
   0.1501440171308133   0.1664009924425898   0.5000008796302637
   0.2905275792049409   0.4972472270730606   0.5000011074140289
   0.7209377640796788   0.4985351370766722   0.4999981997274858
```

```
P-GD-Hex.
   1.00000000000000
     17.0762411269667496    0.0000000000000000    0.0000000000000000
      8.5381205634936137   14.7884586170697698    0.0000000000000000
      0.0000000000000000    0.0000000000000000   20.0000000000000000
   C    P
   18    6
Direct
  0.3883440430082246  0.3883440543888952  0.5000000000000000
  0.4356299059855715  0.4356299211953782  0.5000000000000000
  0.4773877227866876  0.4773877123690156  0.5000000000000000
  0.5226122778704934  0.5226123007386150  0.5000000000000000
  0.5643700946716095  0.5643700919122452  0.5000000000000000
  0.6116559281923060  0.6116559587187282  0.5000000000000000
  0.2233119015188478  0.3883440543888952  0.5000000000000000
  0.3883440468373820  0.2233119111325337  0.5000000000000000
  0.6116559572211600  0.7766880951723678  0.5000000000000000
  0.7766880637903526  0.6116559587187282  0.5000000000000000
  0.5643700904146627  0.8712598287853623  0.5000000000000000
  0.5226122992410325  0.9547754111325588  0.5000000000000000
  0.4356299170452402  0.1287401707168243  0.5000000000000000
  0.4773877082188562  0.0452245883696207  0.5000000000000000
  0.1287422665582625  0.4356299143926563  0.5000000000000000
  0.0452287794621142  0.4773877055662865  0.5000000000000000
  0.8712577055536528  0.5643700851095232  0.5000000000000000
  0.9547711926498081  0.5226122939358930  0.5000000000000000
  0.2698763468212846  0.4602473111648067  0.5000000000000000
  0.4602472997841289  0.2698763543727907  0.5000000000000000
  0.2698763447591048  0.2698763543727907  0.5000000000000000
  0.5397526748177341  0.7301236519321108  0.5000000000000000
  0.7301236538358893  0.5397527019428168  0.5000000000000000
  0.7301236239514495  0.7301236519321108  0.5000000000000000
```

```
P-GD-Rec.
   1.00000000000000
     9.9016546045490905    0.0000000000000000    0.0000000000000000
     0.0000000000000000   16.7570655905802965    0.0000000000000000
     0.0000000000000000    0.0000000000000000   20.0000000000000000
   C    P
   24    4
Direct
  0.1011057073697543  0.0996759729666508  0.5000000000000000
  0.1715359387340740  0.1720553342319917  0.5000000000000000
  0.2409639890193844  0.2332865160045898  0.5000000000000000
  0.3156331645583990  0.2998551476247684  0.5000000000000000
  0.3854864552872002  0.3609163416227830  0.5000000000000000
  0.4561735613616449  0.4333006158390376  0.5000000000000000
  0.9560344620600958  0.0996304136617994  0.5000000000000000
  0.8853592698181743  0.1719518630405403  0.5000000000000000
  0.8159310959978043  0.2331745802463701  0.5000000000000000
  0.7413755650225227  0.2997852240750589  0.5000000000000000
  0.6717749756217017  0.3609307823496479  0.5000000000000000
  0.6012841819335435  0.4333557057693014  0.5000000000000000
  0.4560297745255966  0.5996310039083568  0.5000000000000000
  0.3853614553452758  0.6719521972960862  0.5000000000000000
  0.3159333890186602  0.7331738203378606  0.5000000000000000
  0.2413802865005295  0.7997838472512271  0.5000000000000000
  0.1717687887552444  0.8609282538431913  0.5000000000000000
  0.1012863573383029  0.9333533101445610  0.5000000000000000
  0.6011055244022714  0.5996784488043119  0.5000000000000000
  0.6715345020313563  0.6720563200839038  0.5000000000000000
  0.7409622434022225  0.7332885256840100  0.5000000000000000
  0.8156317951030445  0.7998550946903009  0.5000000000000000
  0.8854878320137871  0.8609164612213718  0.5000000000000000
  0.9561809551118401  0.9332972028209383  0.5000000000000000
  0.2102592774527565  0.0165785241968237  0.5000000000000000
  0.8470914901515840  0.0164415185264630  0.5000000000000000
  0.3470655692375999  0.5164483757043357  0.5000000000000000
  0.7102616211036477  0.5165784477335364  0.5000000000000000
```

```
P-GY-Hex.
   1.00000000000000
     12.6229910288837601    0.0000000000000000    0.0000000000000000
      6.3114955144317930   10.9318309027610496    0.0000000000000000
      0.0000000000000000    0.0000000000000000   20.0000000000000000
   C    P
   12    6
Direct
  0.4068769953922455  0.4077887264159088  0.5000000000000000
  0.4708488244327640  0.4722500769184208  0.5000000000000000
  0.5916282293481316  0.5924481093858631  0.5000000000000000
  0.5272720803400617  0.5283916065036465  0.5000000000000000
  0.1840848011310001  0.4075379814033155  0.5000000000000000
  0.4072386715099725  0.1847145888653614  0.5000000000000000
  0.5919628563713886  0.8152678490382428  0.5000000000000000
  0.8146739385600341  0.5921950533056162  0.5000000000000000
  0.5277258498210742  0.9437157009366359  0.5000000000000000
  0.4714491650206156  0.0562691073688129  0.5000000000000000
  0.0556722166738908  0.4715935633984714  0.5000000000000000
  0.9431075259947548  0.5277367575201311  0.5000000000000000
  0.2468176537094138  0.5048275259978681  0.5000000000000000
  0.5042562568563085  0.2477620427181080  0.5000000000000000
  0.2472036282296060  0.2474710253979353  0.5000000000000000
  0.4946804470731792  0.7525176776653097  0.5000000000000000
  0.7516502826402771  0.4951606832524291  0.5000000000000000
  0.7520251508952356  0.7522250539079280  0.5000000000000000
```

```
P-GY-Rec.
   1.00000000000000
     7.3137354342499892    0.0000000000000000    0.0000000000000000
     0.0000000000000000   12.3515344376282865    0.0000000000000000
     0.0000000000000000    0.0000000000000000   20.0000000000000000
   C    P
  16    4
Direct
  0.1348550644560760  0.1373659798560709  0.5000000000000000
  0.9389822434174633  0.1373511698424394  0.5000000000000000
  0.2353740545448346  0.2343020405173419  0.5000000000000000
  0.8383968685807517  0.2342723630785173  0.5000000000000000
  0.3382980502687047  0.3131770367896962  0.5000000000000000
  0.7356658877880378  0.3132341615161423  0.5000000000000000
  0.4389952145876563  0.4101071227546171  0.5000000000000000
  0.6349542895259077  0.4101333959952100  0.5000000000000000
  0.4389821888442711  0.6373505085721618  0.5000000000000000
  0.6348534307204758  0.6373660491197484  0.5000000000000000
  0.3383972432336009  0.7342725397089325  0.5000000000000000
  0.7353622838369773  0.7343026641163561  0.5000000000000000
  0.2356662625078698  0.8132337069685818  0.5000000000000000
  0.8382972994721456  0.8131798178522942  0.5000000000000000
  0.1349537495950059  0.9101332503120361  0.5000000000000000
  0.9389945578386545  0.9101042200009388  0.5000000000000000
  0.7931695959405332  0.0237729417586081  0.5000000000000000
  0.2807187867022236  0.0237949666050952  0.5000000000000000
  0.2931694837509013  0.5237728125613828  0.5000000000000000
```

```
Al-GD-Hex.
   1.00000000000000
     13.1909355616628208    0.0000000000000000    0.0000000000000000
      6.5954677808314059   11.4236852960858908    0.0000000000000000
      0.0000000000000000    0.0000000000000000   20.0000000000000000
   C    Al
   12     2
Direct
  0.4163193831818788  0.1673483647599809  0.5000000131005322
  0.4703959001774578  0.0591940881052437  0.5000000011250449
  0.4163244879008516  0.4163246158849532  0.5000000022842883
  0.4704010583121416  0.4704021715942730  0.5000000045760018
  0.5295962229974407  0.5295967185057009  0.5000000023231479
  0.5836729962168263  0.5836739418026795  0.4999999903051418
  0.5836701608369381  0.8326496696282462  0.4999999898918404
  0.5295937824946861  0.9408040318116662  0.4999999957134378
  0.1673480265751550  0.4163184540264737  0.5000000204493276
  0.0591941839563148  0.4703954020424490  0.5000000017307471
  0.8326504397857661  0.5836710904234508  0.4999999861182900
  0.9408043777263373  0.5295943561425318  0.4999999916007667
  0.3333316287571364  0.3333326276806901  0.5000000121946897
  0.6666660842628787  0.6666646696848701  0.4999999885867723

  Al-GD-Rec.
   1.00000000000000
     16.6913823920642486    0.0000000000000000    0.0000000000000000
      0.0000000000000000   11.9797222550346518    0.0000000000000000
      0.0000000000000000    0.0000000000000000   20.0000000000000000
   C    Al
   22     2
Direct
  0.4219181421271045  0.4968755553836743  0.4999246596841544
  0.4959299842145555  0.4962142310438438  0.4999955020006936
  0.0788338997709985  0.9973470768461539  0.5000135657885565
  0.5769815696023528  0.4961664718089551  0.5000041197425915
  0.6510001852978036  0.4967302497601196  0.4999734396540205
  0.8219968889840956  0.6342243228719084  0.4999866338689571
  0.8634066001432217  0.7198497880848223  0.4999663343287963
  0.9086977515031762  0.8127741203667114  0.4999658216716156
  0.9503897135472528  0.8974296339823979  0.4999752093003380
  0.8239663951807330  0.3629781382413810  0.5000024302580712
  0.8648448934725721  0.2768155995628376  0.5000097993589705
  0.9092637821454161  0.1829009148210901  0.5000140286133501
  0.9503940087778560  0.0976707685996558  0.5000130042156883
  0.9940959781275112  0.9975129286031574  0.5000009365843709
  0.2505929076811029  0.6338339839315026  0.5000149691215157
  0.2092242367844008  0.7194786766509083  0.5000033238073698
  0.1640143579730733  0.8124625644244290  0.4999833507396829
  0.1223084792095221  0.8970790364064314  0.4999668782468589
  0.2493121958328928  0.3625504885007871  0.5000067475907883
  0.2084686458768132  0.2763296962743240  0.5000358369025051
  0.1640354138962508  0.1824009177090034  0.5000812133775199
  0.1228063813985898  0.0972557341553113  0.5000818648311736
  0.3083615873967176  0.4978026929102484  0.4999876398680740
  0.7645819073581137  0.4978667847395499  0.4999951349009493
```

```
Al-GY-Hex.
   1.00000000000000
     8.7385998001500145    0.0000000000000000    0.0000000000000000
     4.3692999000750072    7.5678494204324851    0.0000000000000000
     0.0000000000000000    0.0000000000000000   20.0000000000000000
   C    Al
   6    2
Direct
  0.4592508384133325  0.0814847470073943  0.5000000398785929
  0.4592029139166840  0.4591992593795240  0.4999999170624889
  0.5407249369640752  0.5407281158476422  0.4999997718193043
  0.5407788986599016  0.9184351162108015  0.4999997995409018
  0.0814853102600708  0.4592521983274338  0.5000004003909169
  0.9184353729581787  0.5407789530123850  0.5000001177757909
  0.3332946044703391  0.3332935751057491  0.5000003021765949
  0.6666285375881387  0.6666295915199285  0.4999996513553810

Al-GY-Rec.
   1.00000000000000
    11.1529645605559509    0.0000000000000000    0.0000000000000000
     0.0000000000000000    7.7685626894044466    0.0000000000000000
     0.0000000000000000    0.0000000000000000   20.0000000000000000
   C    Al
  12    2
Direct
  0.4499939922738108  0.4990567800341452  0.4999823215509451
  0.5606381289406031  0.4987630640948026  0.4999835213199546
  0.8186997027491074  0.7080147209288015  0.5000191346320193
  0.0687631523502148  0.9976849550740923  0.4999986828312615
  0.8773093986876574  0.8421449312582894  0.5000193122154784
  0.8177384804321264  0.2869955303253846  0.5000011805333813
  0.8769994424861025  0.1534468435293235  0.5000013667184078
  0.9428771044048005  0.9979027378997003  0.5000059180591450
  0.1341289499526397  0.8417332550358907  0.4999997685222013
  0.1933778721341852  0.7082260158376883  0.5000001174351638
  0.1936714625405500  0.2869268423776177  0.4999966598797911
  0.1345988294584330  0.1531965743132986  0.4999943256496593
  0.2795925891399662  0.4979291953326239  0.499996286490976
  0.7316107284497875  0.4979785539582693  0.5000015231616146
```

```
As-GD-Hex.
   1.00000000000000
     17.4431125901974795    0.0000000000000000    0.0000000000000000
      8.7215562951091812   15.1061786241503899    0.0000000000000000
      0.0000000000000000    0.0000000000000000   20.0000000000000000
   C    As
   18     6
Direct
  0.3910489357898683  0.3910489471705461  0.5000000000000000
  0.4368419442604932  0.4368419594703070  0.5000000000000000
  0.4779532924598939  0.4779532820422290  0.5000000000000000
  0.5220467081972728  0.5220467310653945  0.5000000000000000
  0.5631580563966807  0.5631580536373164  0.5000000000000000
  0.6089510354106551  0.6089510659370774  0.5000000000000000
  0.2179021159556100  0.3910489471705461  0.5000000000000000
  0.3910489396190329  0.2179021255692959  0.5000000000000000
  0.6089510644395091  0.7820978807356056  0.5000000000000000
  0.7820978493535904  0.6089510659370774  0.5000000000000000
  0.5631580521397339  0.8736839053351844  0.5000000000000000
  0.5220467295678120  0.9559065504789857  0.5000000000000000
  0.4368419553201619  0.1263160941669952  0.5000000000000000
  0.4779532778920697  0.0440934490231939  0.5000000000000000
  0.1263181900084334  0.4368419526675851  0.5000000000000000
  0.0440976401156874  0.4779532752395070  0.5000000000000000
  0.8736817821034748  0.5631580468345945  0.5000000000000000
  0.9559023319962350  0.5220467242626725  0.5000000000000000
  0.2661038370196209  0.4677923307681482  0.5000000000000000
  0.4677923193874776  0.2661038445711270  0.5000000000000000
  0.2661038349574412  0.2661038445711270  0.5000000000000000
  0.5322076552143926  0.7338961617337745  0.5000000000000000
  0.7338961636375529  0.5322076823394752  0.5000000000000000
  0.7338961337531131  0.7338961617337745  0.5000000000000000
```

```
As-GD-Rec.
   1.00000000000000
     10.0916634603880642    0.0000000000000000    0.0000000000000000
      0.0000000000000000   17.0110213075002790    0.0000000000000000
      0.0000000000000000    0.0000000000000000   20.0000000000000000
   C    As
    24    4
Direct
  0.0991195475620543  0.1044644167345012  0.5000000000000000
  0.1702681748609862  0.1747051167527829  0.5000000000000000
  0.2413438929285832  0.2339704709958070  0.5000000000000000
  0.3166946359900251  0.2986313461103265  0.5000000000000000
  0.3869663897236819  0.3582314515281055  0.5000000000000000
  0.4576714421416312  0.4285863852871401  0.5000000000000000
  0.9576953201756737  0.1044747610529271  0.5000000000000000
  0.8868496936336285  0.1748173947668192  0.5000000000000000
  0.8163826849392564  0.2343339686139458  0.5000000000000000
  0.7409330581315103  0.2989312559314072  0.5000000000000000
  0.6700401978831678  0.3582629599565905  0.5000000000000000
  0.5991004559386965  0.4285859819191202  0.5000000000000000
  0.4576808617125607  0.6044734582954661  0.5000000000000000
  0.3868329202527505  0.6747961486132894  0.5000000000000000
  0.3163648521629980  0.7343241660292392  0.5000000000000000
  0.2409171227182156  0.7989242712928046  0.5000000000000000
  0.1700278300077187  0.8582773527767813  0.5000000000000000
  0.0991128471652161  0.9286020036029612  0.5000000000000000
  0.5991362994748997  0.6044673060339534  0.5000000000000000
  0.6702657812147166  0.6747114582339222  0.5000000000000000
  0.7413443035205987  0.7339705076730496  0.5000000000000000
  0.8166955454890683  0.7986233435897816  0.5000000000000000
  0.8869676483798514  0.8582277738569388  0.5000000000000000
  0.9577022820420424  0.9285917846802718  0.5000000000000000
  0.2163952620891223  0.0165103961462592  0.5000000000000000
  0.8405760800537507  0.0164467282225829  0.5000000000000000
  0.3405767734927494  0.5164475812064069  0.5000000000000000
  0.7163373245927147  0.5164900597765580  0.5000000000000000
```

```
As-GY-Hex.
   1.00000000000000
     12.9938772976492896    0.0000000000000000    0.0000000000000000
      6.4969386488142700   11.2530278334269500    0.0000000000000000
      0.0000000000000000    0.0000000000000000   20.0000000000000000
   C    As
   12    6
Direct
  0.4086025158527136  0.4123646835437427  0.5000000000000000
  0.4696076181900679  0.4747210283805927  0.5000000000000000
  0.5869749856656483  0.5908098138874749  0.5000000000000000
  0.5245822655128549  0.5298673047348359  0.5000000000000000
  0.1777931722900448  0.4091922957249352  0.5000000000000000
  0.4117488727315148  0.1783028623873122  0.5000000000000000
  0.5902276200343692  0.8216354777034596  0.5000000000000000
  0.8209911381259971  0.5876339312557803  0.5000000000000000
  0.5291613952247260  0.9449276588628734  0.5000000000000000
  0.4740591000504040  0.0550108820137609  0.5000000000000000
  0.0544653341174595  0.4701358810706324  0.5000000000000000
  0.9443476435161102  0.5252485214640075  0.5000000000000000
  0.2408688199638505  0.5134768145436652  0.5000000000000000
  0.5128621418511727  0.2449950910584207  0.5000000000000000
  0.2445430251162435  0.2413377896080888  0.5000000000000000
  0.4859425716341335  0.7587010057223509  0.5000000000000000
  0.7543449449956867  0.4865301120564212  0.5000000000000000
  0.7580514091270132  0.7549819759816572  0.5000000000000000
```

```
As-GY-Rec.
   1.00000000000000
     7.4389940700838864    0.0000000000000000    0.0000000000000000
     0.0000000000000000   12.6650232932029194    0.0000000000000000
     0.0000000000000000    0.0000000000000000   20.0000000000000000
   C    As
   16    4
Direct
  0.1324521476419847  0.1431930693205956  0.5000000000000000
  0.9410332511764565  0.1431465625538308  0.5000000000000000
  0.2341123612682168  0.2361151848375087  0.5000000000000000
  0.8396876111876779  0.2361626082608552  0.5000000000000000
  0.3399161564391449  0.3112571878502592  0.5000000000000000
  0.7343126444704211  0.3114929015012322  0.5000000000000000
  0.4413269882397657  0.4042096086461839  0.5000000000000000
  0.6327285571624586  0.4043892959858582  0.5000000000000000
  0.4410330184963414  0.6431458031931498  0.5000000000000000
  0.6324561050586794  0.6431914900719775  0.5000000000000000
  0.3396868097158148  0.7361578450737483  0.5000000000000000
  0.7341127342999627  0.7361174086527527  0.5000000000000000
  0.2343126430347198  0.8114904792305992  0.5000000000000000
  0.8399178494417114  0.8112576626435555  0.5000000000000000
  0.1327252805180308  0.9043948769906009  0.5000000000000000
  0.9413273580116552  0.9042142995428577  0.5000000000000000
  0.7848111101182482  0.0236098669572087  0.5000000000000000
  0.2890201876080596  0.0239364855688038  0.5000000000000000
  0.2848092919273171  0.5236087047319984  0.5000000000000000
  0.7890232743246770  0.5239306049195065  0.5000000000000000
```

```
   Ga-GD-Hex.
      1.00000000000000
        13.2503228556836792    0.0000000000000000    0.0000000000000000
         6.6251614278418387   11.4751162013699908    0.0000000000000000
         0.0000000000000000    0.0000000000000000   20.0000000000000000
     C    Ga
      12    2
Direct
   0.4168211591972693  0.1663465669688975  0.5000000054087863
   0.4705046454797071  0.0589776716865273  0.5000000022487967
   0.4168250928640873  0.4168248954147842  0.5000000057090332
   0.4705090270905643  0.4705101506542846  0.5000000017424000
   0.5294878528397078  0.5294887678875213  0.5000000015687718
   0.5831717445732139  0.5831735021044793  0.5000000047087312
   0.5831685518761702  0.8336511215681313  0.4999999892466818
   0.5294860254475822  0.9410198900390085  0.4999999960918800
   0.1663466580220287  0.4168198796362290  0.5000000160412839
   0.0589781916718906  0.4705039377429614  0.5000000023367903
   0.8336518797702226  0.5831699183291832  0.4999999841034821
   0.9410206790657654  0.5294866241615992  0.4999999907135262
   0.3333313685768218  0.3333325302040393  0.5000000096051238
   0.6666658567067998  0.6666647456955275  0.4999999904747554

   Ga-GD-Rec.
      1.00000000000000
        16.6540885105193013    0.0000000000000000    0.0000000000000000
         0.0000000000000000   12.0675236196701299    0.0000000000000000
         0.0000000000000000    0.0000000000000000   20.0000000000000000
     C    Ga
      22    2
Direct
   0.4218459260096168  0.4964931504194041  0.5000226827920358
   0.4958275883804006  0.4964673349217605  0.5000215655362084
   0.0789816888856620  0.9971833195339386  0.4999961682955529
   0.5771315882547441  0.4965044390364071  0.5000390814215692
   0.6511168935797969  0.4965294783132563  0.5000165716264604
   0.8240185078533955  0.6347160811294970  0.4999989186032607
   0.8646532021411488  0.7202284466412223  0.4999996698422322
   0.9090358766711830  0.8134762995806426  0.4999993979899529
   0.9502693973440230  0.8980086228857971  0.5000000787519525
   0.8261751797291552  0.3622530321975574  0.5000056995849320
   0.8665366662608065  0.2764663062768733  0.4999958325842186
   0.9101749492265938  0.1824659221911773  0.4999757846075497
   0.9505883193566333  0.0971483990462687  0.4999716135889685
   0.9940118693868811  0.9974183528205529  0.4998870095017300
   0.2488879649328339  0.6344038659796780  0.5000537980226483
   0.2082087365176051  0.7198734413931192  0.5000362513688046
   0.1637182643337454  0.8130076748342887  0.5000271810368204
   0.1225292414691452  0.8975852398041013  0.5000059941804977
   0.2467864883800530  0.3620569130227480  0.4999150117810913
   0.2064376659872593  0.2762665040438606  0.4999714547946041
   0.1628736325069511  0.1821933193276877  0.4998805720783712
   0.1226359113741040  0.0967144473722783  0.4999819525359470
   0.3068757714289490  0.4974282369611629  0.4999863457717311
   0.7661045762914824  0.4976615479460094  0.5000138081594869
```

```
Ga-GY-Hex.
   1.00000000000000
     8.7947210559296067    0.0000000000000000    0.0000000000000000
     4.3973605279648016    7.6164518536298784    0.0000000000000000
     0.0000000000000000    0.0000000000000000   20.0000000000000000
   C   Ga
    6    2
Direct
  0.4596574922302636   0.0806688379379565   0.4999999945189799
  0.4596076288714670   0.4596033510928095   0.4999998958912002
  0.5403132687911238   0.5403151394530994   0.4999998946785453
  0.5403712724472527   0.9192522158431089   0.4999998966410928
  0.0806694991404839   0.4596579110833829   0.5000002736979923
  0.9192533362950002   0.5403727920773846   0.5000001307869013
  0.3332976713742539   0.3332967404419165   0.5000003062133942
  0.6666312440808682   0.6666345684811503   0.4999996075718585

Ga-GY-Rec.
   1.00000000000000
    11.2015821655039396    0.0000000000000000    0.0000000000000000
     0.0000000000000000    7.7994668115346819    0.0000000000000000
     0.0000000000000000    0.0000000000000000   20.0000000000000000
   C   Ga
   12    2
Direct
  0.4505644876204968   0.4989732516911403   0.4999976680634788
  0.5603340009172157   0.4989110397850141   0.4999977891951275
  0.8195793412323056   0.7095869489764652   0.4999985307655166
  0.0684659713829987   0.9977183913590721   0.4999994435192008
  0.8781826066088954   0.8425882929299675   0.4999988842436025
  0.8193017075886289   0.2862557960404502   0.4999957111172577
  0.8778843887770265   0.1532336699588939   0.4999888301208699
  0.9432091143948398   0.9980113574125500   0.4999986185136933
  0.1331991189264699   0.8420728971445115   0.5000051759803412
  0.1914805243586315   0.7088347913533113   0.5000243159819604
  0.1927867229558444   0.2854335655204494   0.4999980119019867
  0.1339489524535082   0.1526854667638275   0.4999975500078477
  0.2790942426861491   0.4976690263160890   0.5000039577133819
  0.7319686540969670   0.4980357264195518   0.4999956318757199
```

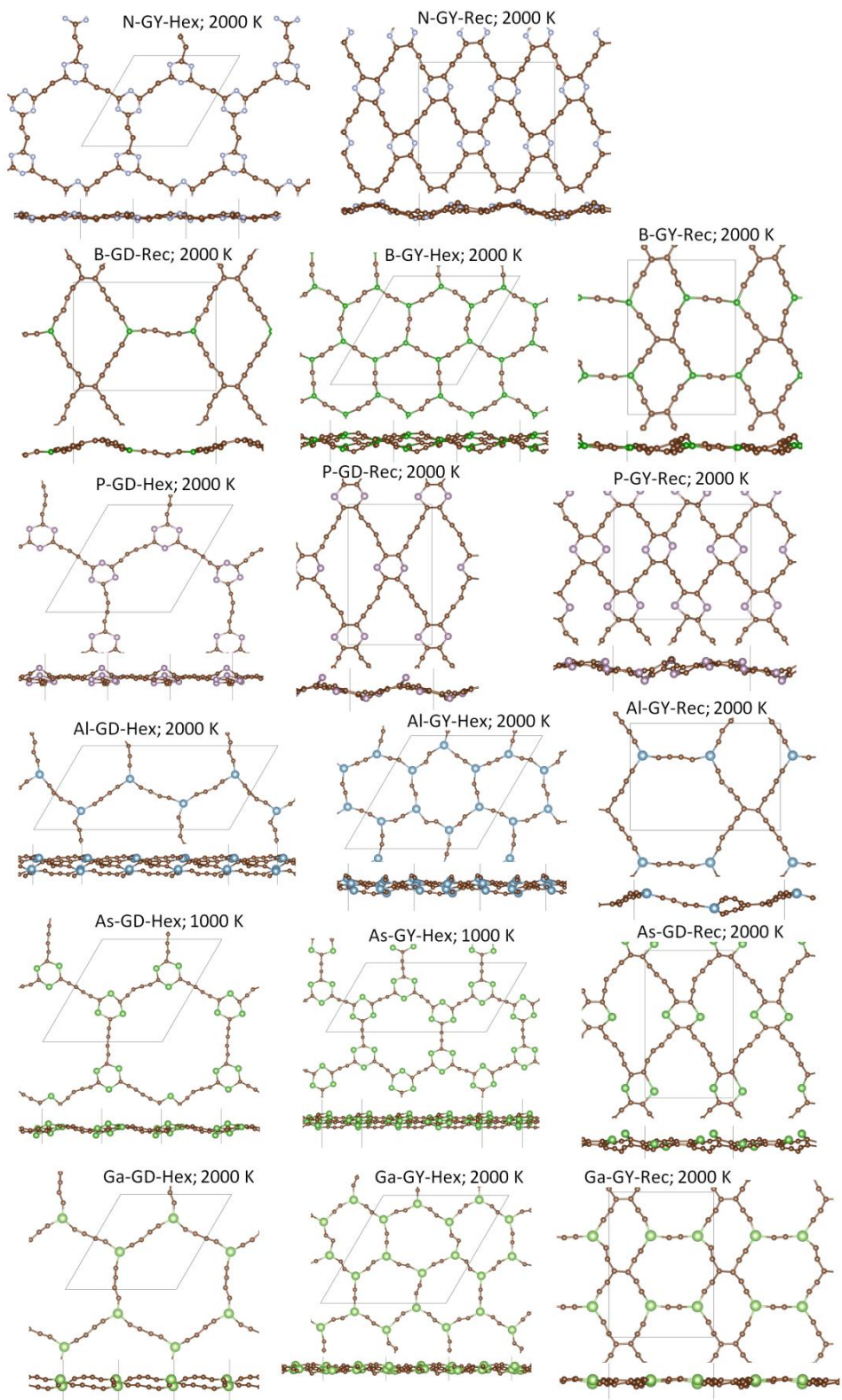

**Fig. S1**, Top and side views of predicted monolayers at different temperatures after the AIMD simulations for 20 ps. The black lines define the simulated systems.

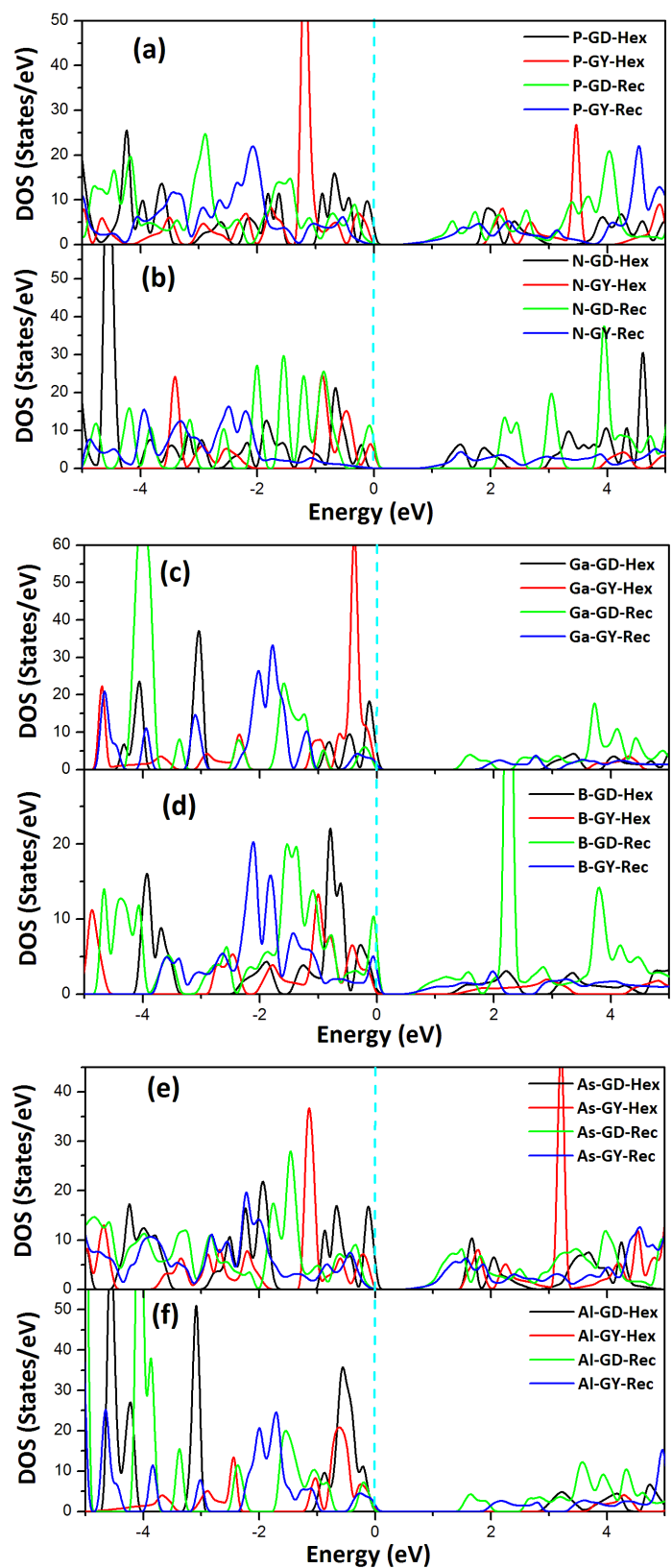

**Fig. S2**. Total electronic density of states predicted by the HSE06 functional. The Fermi energy is aligned to zero.